\documentclass[aps,prb]{revtex4}
\usepackage{graphicx}
\usepackage{amsmath}
\usepackage{bm}
\usepackage{color}
\usepackage{hyperref} 

\begin{document}
\title{Enhancement of Zener tunneling rate via electron-hole attraction within a time-dependent quasi-Hartree-Fock method}

\author{Yasushi Shinohara}
\affiliation{NTT Basic Research Laboratories, NTT Corporation, 3-1 Morinosato Wakamiya, Atsugi, Kanagawa 243-0198, Japan}
\author{Haruki Sanada}
\affiliation{NTT Basic Research Laboratories, NTT Corporation, 3-1 Morinosato Wakamiya, Atsugi, Kanagawa 243-0198, Japan}
\author{Katsuya Oguri}
\affiliation{NTT Basic Research Laboratories, NTT Corporation, 3-1 Morinosato Wakamiya, Atsugi, Kanagawa 243-0198, Japan}

\date{\today}
\begin{abstract}
The tunneling process, a prototypical phenomenon of nonperturbative dynamics, is a natural consequence of photocarrier generation in materials irradiated by a strong laser. Common treatments for Zener tunneling are based on a one-body problem with a field-free electronic structure. In a literature Ref. \onlinecite{Ikemachi2018}, a characteristic of gap shrinking or excitation can occur due to the electron-hole interaction for slow and strong time-varying electric fields. We have developed a theoretical framework called the quasi-Hartree-Fock (qHF) method to enable a more flexible imitation of the electronic structures and electron-hole attraction strength of materials compared to the original Hartree-Fock method. In the qHF framework, band gap, reduced effective mass, and electron-hole interaction strength can be independently selected to reproduce common crystals. In this study, we investigate the effect of electron-hole attraction on Zener tunneling subjected to a DC electric field for four different systems using the qHF method. Our findings demonstrate that the electron-hole attraction promotes the tunneling rates in all four material systems assumed as examples. Specifically, systems that have a strong electron-hole interaction show a few factor enhancements for tunneling rates under DC fields, while systems with a weak interaction show higher enhancements of a few tens of percent.
\end{abstract}
\maketitle

\section{Introduction}
Recent progress with strong coherent light source technology has enabled the application of a transient V/nm scale strong field to materials only on the femtosecond time scale. Materials under a strong and short light field exhibit extremely nonlinear responses to the field \cite{Rossi2002, Ghimire2014, Kruchinin2018, Ghimire2019, Yue2022} because the field strength is comparable to the electric fields felt by the valence electrons in the materials. Such a strong light field excites photocarriers even in insulators with off-resonant photons, which undergo ablation and permanent damage due to their nonlinear responses \cite{Chichkov1996, Balling2013, Sugioka2014}. When a strong field is applied to vapor, it results in one of the most symbolic phenomena, namely, high-order harmonic generation, in which three steps (field ionization of an electron, electron acceleration by the field, and recombination to the parent ion) govern the mechanism \cite{Corkum1983}. The number of excited electrons is one of the most important physical quantities of the material exposed to a strong field. According to Keldysh's theory \cite{Keldysh}, when a strong field makes the tunneling ionization faster than the time scale of the electric field oscillation, i.e., the Keldysh parameter $\gamma < 1$, the optical absorption rate is explained by the ionization due to tunneling rather than that due to multi-photon excitation.

The correlation between the excited electron and the hole corresponds to a local electric field of the same order of magnitude as that of a strong optical pulse. The electron-hole (e-h) interactions are pronounced by the formation of excitons, which play an important role in linear optical absorption processes, especially in systems with low dielectric constants, low-dimensional structures, and at low temperatures. Material-dependent excitons are well described by the Bethe-Salpeter equation \cite{Onida2002, Leng2016, Reining2018} through the dielectric function based on an atomistic quantum mechanical description. The size of the exciton ranges from the sub-nm level to a few nm and its biding energy is from meV to sub-eV. The corresponding field strength can be up to the order of V/nm for tightly bound excitons. While this field strength is comparable to the typical strength of nonlinear optics, the role of exciton existence in nonlinear optical excitation has not yet been investigated.

In rare cases, however, the effects of e-h interaction on the Zener tunneling process \cite{Zener1934} have been investigated.
The field-induced tunneling property of electrons has been studied mainly in quantum wells \cite{Carlo1994, Sibille1998, Rosam2001}, where theoretical treatments are based on the assumption of rigid energy bands.
The Zener tunneling dynamics of atoms in optical lattices also have been investigated mainly within independent particle approximation \cite{Bharucha1997, Glueck1999, Holthaus2000, Takahashi2017}.
A mean-field treatment of the interatomic potential has been reported \cite{Wu2000}, but no direct clue was obtained for the e-h effect on the Zener tunneling because the focus was on Bosonic systems and contact interaction rather than Fermionic systems and long-range interaction.
The semiconductor Bloch equation (SBE) \cite{HaugKoch, Lindberg1988} and time-dependent Hartree-Fock (TD-HF) are the theoretical frameworks to address the effects of e-h on the Zener tunneling.
For example, Garg et al. reported the prominent enhancement of photoradiation intensity (i.e., the power spectrum of dipole acceleration) by increasing the e-h interaction strength through SBE simulations \cite{Garg2016}.
Ikemachi et al. reported a similar enhancement for photoradiation by means of TD-HF calculations for a one-dimensional model \cite{Ikemachi2018}.
Both investigations did not make an explicit argument for the number of excited electrons but photoradiation enhancement by e-h interaction.

Parts of the study by Ikemachi et al. \cite{Ikemachi2018} indicated that the e-h interaction promotes e-h pair creation, although direct numerical proof was not given.
To clarify the background of our current investigation, we briefly review the related part of the study.
In their work, TD-HF and {\it frozen} TD-HF, where the one-body Hamiltonian is fixed to the initial self-consistent solution, are introduced to calculate high-order harmonic generation for a pulsed electric field.
Time-frequency analysis of frozen TD-HF shows good agreement between the interband frequency of an e-h pair and transient spectra, where e-h pairs are created around the point where the energy difference between the conduction and the valence band is minimum in the Brillouin zone (BZ).
The spectrum calculated by the full TD-HF incorporating e-h interactions shows a higher frequency emission compared to that obtained by the frozen TD-HF.
This higher frequency component invoking the e-h pair creation is promoted by the e-h interaction.
In the literature, a {\it hauling-up effect} was proposed to explain the promotion due to the e-h interaction.
When the representation that a Houston function and its coefficient for the orbital function is used, the finite component of the e-h pair and the off-diagonal component of the density-matrix in a $k$-point behave as a resonant field to other $k$-points.
This effect explains why the e-h pair creation is promoted even far from the narrowest point in the gap in the BZ thanks to the e-h pair creation around the gap's narrowest point, as described in Fig. 5 of Ikemachi et al.’s paper \cite{Ikemachi2018}.

While TD-HF is a concrete theoretical foundation of the Fermionic many-body theory, it is unsuitable for simulating realistic materials' non-perturbative dynamics.
A fully converged real-time solution, including a long-range exchange interaction for a given periodic potential, is still limited to spatially one-dimensional systems \cite{Ikemachi2018, Williams2021} because of the high computational cost.
The expected electronic structures to imitate common solids (e.g., GaAs and $\alpha$-quartz) are difficult to model by choosing a one-body external potential within the TD-HF framework.
Furthermore, TD-HF cannot reproduce the realistic exciton binding energy of such common materials since the bare e-h interaction potential in TD-HF is too strong.
The electronic structure is not only affected by the one-body external potential but also by the e-h interaction potential stemming from the self-consistent field manner.
Therefore, most of the TD-HF simulations based on the spatial grid representation have been limited to artificial or quite simple systems such as spatially one-dimensional atomic arrays, atoms, or small molecules that contain a few valence electrons.

In light of this background, we have developed a theoretical framework that enables the flexible incorporation of e-h interactions for electron quantum dynamics under a time-dependent external field.
The e-h interaction is included in a mean-field level in a quasi-Hartree-Fock (qHF) treatment on top of the electronic structure of an independent electron system (IES).
Our qHF provides a simple protocol to obtain a band gap and reduced effective mass values for IES by choosing a one-body potential.
The strength of the e-h interaction is independently selected to reproduce an expected exciton binding energy without affecting the IES in the qHF framework.
We perform explicit time evolution of the quantum system under a time-dependent electric field while assuming a DC field as the electric field that induces the tunneling process.
This assumption is reasonable when $\gamma$ is sufficiently smaller than 1, such as the condition ($\gamma = 0.3$ with 2.5 V/nm) discussed in Ikemachi et al.'s research \cite{Ikemachi2018}.
The tunneling rates are estimated from the increasing rate of the number of excited electrons obtained by the time-dependent simulation.
We investigate the tunneling rate of electrons under a strong electric field, particularly focusing on the dependence on the electronic structures and the e-h interaction strength.

Section \ref{section:theoretial_framework} of this paper presents our theoretical framework.
Specifically, we introduce a time-dependent mean-field theoretical framework for the system consisting of interacting electrons, in which we can flexibly set an arbitrary band gap and an effective mass that determine quasi particle spectra.
In Sec. \ref{section:simulation_scheme}, we introduce the details of the simulation protocol, including the complete protocols for numerical simulations and the connection of our tunneling problem.
Section \ref{section:results} shows the results of the simulation assuming four different material systems.
In Sec. \ref{section:discussion}, we present the discussion, and conclusions are given in Sec. \ref{section:conclusion}.
The atomic unit is used throughout this paper, where the elementary charge $e$, electron mass $m_e$, and Dirac constant $\hbar$ are set to 1.

\section{Theoretical framework}
\label{section:theoretial_framework}
We derive the equation of motion of the time-dependent quasi-Hartree-Fock (TD-qHF) method from the original TD-HF method.
An arbitrary one-body potential can be chosen for the ground state electronic structure within the TD-qHF framework.
The potential is determined such that the electronic structure, precisely band gap and reduced effective mass, imitates common materials.
The strength of the e-h interaction is independently tuned to give the correct exciton binding energy of the same material.
This independent determination of the electronic structure and exciton binding energy is not realized in the original TD-HF framework.

We present the theoretical foundation of the TD-qHF method along three types of properties: (a) the interaction among electrons does not change the electronic ground state as long as the applied field is zero, (b) the energy to be conserved without the applied field can be defined in terms of orbital functions, and (c) an explicit orbital set that satisfies the conditions of energy minima. These inspections manifest that our initial wave function is the stable point of the total energy and has almost the same theoretical foundation as TD-HF.

\subsection{Time-dependent quasi-Hartree-Fock method}
We begin with the spatially one-dimensional TD-HF equation based on the Born-von-Karman (BvK) boundary condition:
\begin{align}
 \mathrm{i} \frac{\partial}{\partial t}\psi_{i,k}(x,t)
 = 
 \left[\frac{1}{2}\left( -\mathrm{i}\frac{\partial}{\partial x} + A(t) \right)^2 + v_{\mathrm{ext}}(x) \right]\psi_{i,k}(x,t) \notag \\
 +\hat{v}_{\mathrm{MF}}[\rho]\psi_{i,k}(x,t) \label{eq:TD-HF},\\
 \hat{v}_{\mathrm{MF}}[\rho]\psi_{i,k}(x,t)
 = \int_0^{N_k a} \! \mathrm{d}x' \rho(x',x',t) w(x -x') \psi_{i,k}(x,t)\notag \\
 -\frac{1}{2} \int_0^{N_k a} \! \mathrm{d}x' \rho(x,x',t) w(x -x' )\psi_{i,k}(x',t),\label{eq:TD-HF-MFpart} \\
 \rho(x,x';t)
 =
 2\sum_{i,k} \psi_{i,k}(x,t) \psi_{i,k}^*(x',t),\\
 \psi_{i,k}(x,t)
 =
 \frac{1}{\sqrt{N_k}}e^{ikx}u_{ik}(x,t),
\end{align}
where $a$, $v_{\mathrm{ext}}$, $w(x-x')$, and $N_k$ are the lattice constant, the external potential, the modeled Coulombic interaction between electrons, and the number of primitive cells for the simulation box.
The summation in the density-matrix $\rho$ is taken over occupied bands.
The number of primitive cells $N_k$ should be as high as possible to obtain well converged results. The external potential has the lattice periodicity $v_{\mathrm{ext}}(x+a) = v_{\mathrm{ext}}(x)$. The crystal momentum $k$ has the discretized values of $0, 2\pi/(a N_k), 4\pi/(a N_k), \dots, 2(N_k -1)\pi/(a N_k)$. The external potential is supposed to be the Coulombic attraction potential from the ion array. The orbital function $\psi_k$ is orthonormalized in the simulation cell as $ \int_0^{N_k a}\! \mathrm{d}x \ \psi_{i,k}^*(x,t) \psi_{j,k'}(x,t) =\delta_{i,j}\delta_{k,k'}$, while the lattice periodic part (LPP) $u_{ik}$ is orthonormalized in the unit cell. Here, the spin-restricted wave function for the Slater determinant is assumed, i.e., up and down spin orbitals are the same spatial orbital. A $k$-resolved density-matrix is a useful intermediate variable, as 
\begin{align}
 \rho^k(x,x';t) = 2\sum_{i} u_{ik}(x,t)u_{ik}^*(x',t),
\end{align}
with $\rho(t) = \frac{1}{N_k}\sum_k e^{\mathrm{i}kx}\rho^k(t)e^{-\mathrm{i}kx}$.

The mean-field part \eqref{eq:TD-HF-MFpart} in the TD-HF equation has a linear dependence on the density matrix $\rho$. For an arbitrary density matrix $\rho_0$, the mean-field part has the following relation:
\begin{align} 
 \hat{v}_{\mathrm{MF}}[\rho]\psi_{i,k}(x,t)
 =
 \hat{v}_{\mathrm{MF}}[\rho_0]\psi_{i,k}(x,t)
 +
 \hat{v}_{\mathrm{MF}}[\rho - \rho_0]\psi_{i,k}(x,t). 
\end{align}
Now, we introduce an approximation to the treatment of the Hamiltonian, as 
\begin{align} 
 \left\{ v_{\mathrm{ext}}(x) + \hat{v}_{\mathrm{MF}}\left[\rho_0 \right]\right\}\psi_{i,k}(x,t)
 \to
 v(x)\psi_{i,k}(x,t) ,\\
 \rho_0(x,x';t)
 =
 e^{+\mathrm{i}A(t)x} \left( 2\sum_{i,k} \phi_{i,k}(x)\phi_{i,k}^*(x') \right) e^{-\mathrm{i}A(t)x'},\\
 \left[\frac{1}{2}\left( -\mathrm{i}\frac{\partial}{\partial x} + k \right)^2 + v(x) \right]v_{i,k}(x)
 =
 \epsilon_{i,k}v_{i,k}(x),\\
 \phi_{i,k}(x) = \frac{1}{\sqrt{N_k}} e^{\mathrm{i}kx} v_{ik} (x),\\
 \psi_{i,k}(x,t=0) = \phi_{i,k}(x),
\end{align}
where the effective one-body potential is assumed to be the same lattice periodicity $v(x+a)=v(x)$. Note that $\rho_0$ parametrically depends on the time due to the velocity gauge choice. The mean-field potential at the Hartree-Fock ground state is replaced by the spatially local potential $v$. We solve the following time-dependent equation of motion rather than the original time-dependent HF \eqref{eq:TD-HF}:
\begin{align}
 \mathrm{i} \frac{\partial}{\partial t}\psi_{i,k}(x,t)
 = 
 \left[\frac{1}{2}\left( -\mathrm{i}\frac{\partial}{\partial x} + A(t) \right)^2 + v(x) \right]\psi_{i,k}(x,t) \notag \\
 +\hat{v}_{\mathrm{MF}}[\rho - \rho_0]\psi_{i,k}(x,t), \label{eq:TD-qHF}
\end{align}
called the TD-qHF equation.

The beauty of TD-qHF is that we can reduce it to an equation of motion for IES by omitting the MF part from Eq. \eqref{eq:EoM-LPP-TD-qHF}. The dynamics of IES, as determined by the Hamiltonian $\left[-\mathrm{i}\partial/\partial x+A(t)\right]^2/2+v(x)$, are regarded as reference dynamics. We can solely change the influence of the electron-hole interaction on the dynamics by changing the strength of the MF term. We can also prepare an arbitrary potential for $v$ to create an expected electronic structure to simulate different materials, from ideal toy crystals to common crystals. This helpful nature is impossible for the original TD-HF because the modeled Coulombic interaction $w$ affects the dynamics as well as the quasi-particle spectra of the reference system.

Here, let us explain the relation between TD-qHF and the famous SBE \cite{HaugKoch, Lindberg1988}. TD-qHF equation is equivalent to SBE when the LPP eigenfunction of $-\tfrac{1}{2}\tfrac{\partial^2}{\partial x^2} + v(x)$ does not depend on $k$ and the short-range part of the Coulombic potential is neglected (see Appendix \ref{section:appendixB}). TD-qHF is therefore applicable to systems that have tightly bound excitons (e.g., Frenkel exciton). Coulomb matrix elements between pairwise orbitals are more naturally introduced than when using SBE. Moreover, TD-qHF via Eq. \eqref{eq:TD-qHF} is free from the phase determination of the dipole matrix element \cite{Li2019, Jiang2020} unless the Houston basis is introduced.

\begin{figure}[h!]
 \centering
 \includegraphics[width=1.0\linewidth]{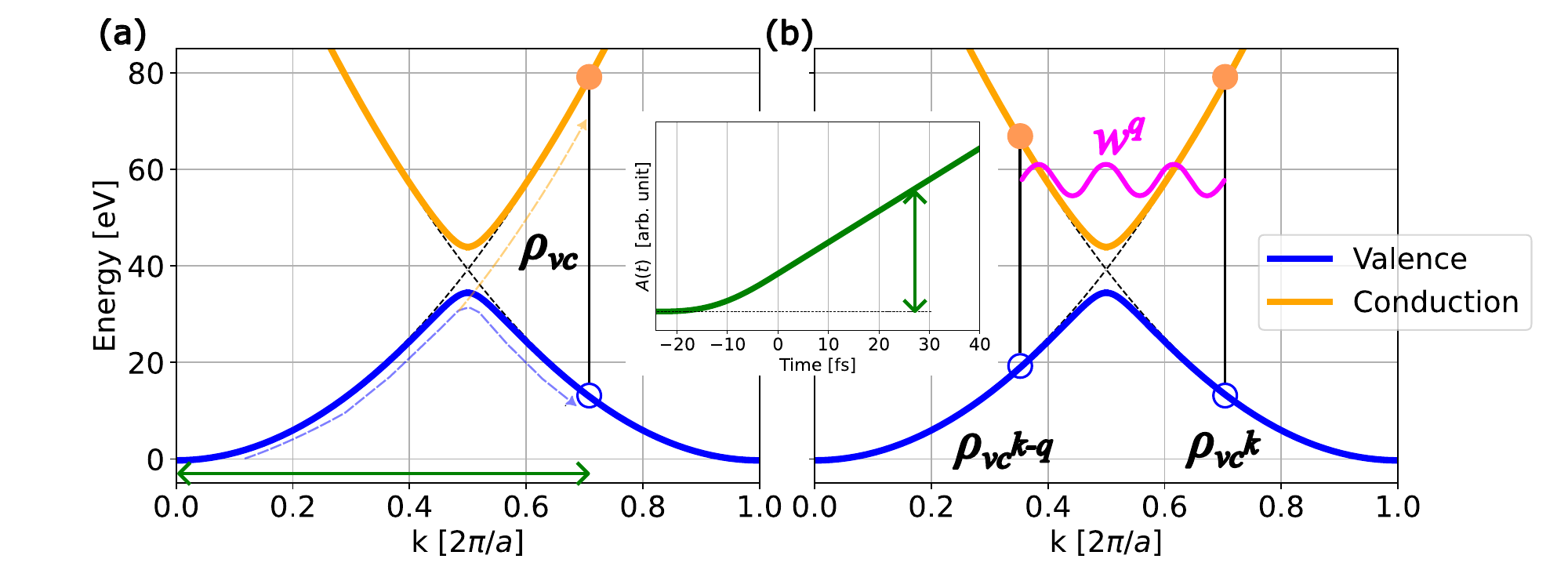}
 \caption{ Schematic image of tunneling ionization for (a) IES and (b) TD-qHF on top of IES. The band structure for a Hamiltonian of 1DTDHF as an example is denoted by solid curves. The $A$-field shape is shown in the inset. Nonadiabatic transition occurs when an orbital in $k$-space passes the narrowest point of the gap due to the intraband motion via the $A$-field, where diabatic levels are drawn as dashed curves. }
 \label{fig:Schematic}
\end{figure}

A representation of the physical picture embedded in TD-qHF is depicted in Fig. \ref{fig:Schematic}.
All $k$-points of valence (conduction) bands are fully occupied (unoccupied) initially because our system is an band insulator.
The off-diagonal component of the density matrix $\rho_{cv} = \left\langle v_{v,k+A(t)} \left| \rho^k \right| v_{c,k+A(t)} \right \rangle $ appears when an e-h pair is created.
The Landau-Zener (LZ) model provides a good description when only two diabatic components dominate the electronic structure around the gap narrowest point (GNP) ($k=\pi/a$ in this case), as discussed in Sect. \ref{section:gap_mass_LZmodel}, where $k+A(t)$ is the time-dependent parameter of the LZ model.
Through this process, the population in the conduction (valence) band increases (decreases), as indicated by the filled disk (open circle) in Fig. \ref{fig:Schematic}(a).
The e-h pairs on different $ k$ points are totally independent of IES by definition. When a dynamic started from a $\Gamma$-point is drawn, different $k$-points reveal the same dynamic.
This bifurcation process starting from a $k$-point alternately happens in time because an $A(t) \sim -E_{\mathrm{DC}}t$ shape is assumed for the DC field, and the Brillouin zone has periodicity in the reciprocal space. By including an e-h interaction, density matrix components in different $ k$ points interact with each other by means of the Coulomb interaction $w$.
When we consider only the long-range part of the Coulomb potential written as $w^q$, an additional field $-\sum_q w^q \rho_{vc}^{k-q}$ appears in the equation of motion for $\rho_{vc}^k$, as described by Eq. \eqref{eq:EoM-DM-SBE}.
Note that the Coulomb potential $w$ in TD-qHF not only couples $\rho_{vc}^k$ and $\rho_{vc}^{k-q}$ but also other density matrix components, as discussed in Sect. \ref{section:simulation_scheme} and Appendix \ref{section:appendixB}.

\subsection{Theoretical foundation of TD-qHF method}
We need to examine this equation of motion \eqref{eq:TD-qHF} to determine whether or not the expected conditions are satisfied. The first point is to see whether the arbitrary one-body observable $\sum_{ik} \left\langle \psi_{ik} (t) \left| \hat{o} \right| \psi_{ik} (t)\right\rangle$ is a constant in time without the field. This condition is confirmed by the fact that a trial solution $\tilde{\psi}_{i,k}(x,t)=e^{-\mathrm{i}\epsilon_{i,k}t}\phi_{i,k}(x)$ is the solution of Eq. \eqref{eq:TD-qHF} without the field, since $\hat{v}_{\mathrm{MF}}$ vanishes when $\rho$ from the trial solution is equivalent to $\rho_0$. The solution does not change observables of any one-body operator. Therefore, $\hat{v}_{\mathrm{MF}}$ does not change the Slater determinant composed by $\psi_{i,k}$ as long as we set the proper initial condition for $\psi_{ik}$ when the field is absent.

For the second point, we need to see if there is an energy expression as a function of $\psi_{ik}$. This energy expression should give a constant value for an arbitrary $\psi_{ik}$, not necessarily $\phi_{i,k}$ at initial, obeying Eq. \eqref{eq:TD-qHF} as long as the external field is zero. A possible energy functional is given as
\begin{align} 
 E[\psi, \psi^*]
 =
 2\sum_{i,k}\int \! \mathrm{d}x \notag \\
 \psi_{i,k}^* (x,t) \left[\frac{1}{2}\left( -\mathrm{i}\frac{\partial}{\partial x} + A(t) \right)^2 + v(x) \right]\psi_{i,k}(x,t) \notag \\
 +
 \sum_{i,k} \int \! \mathrm{d}x \left[ \psi_{i,k}^* (x,t) \hat{v}_{\mathrm{MF}}[\rho - \rho_0]\psi_{i,k}(x,t) \right. \notag \\
 \left. - e^{-\mathrm{i}A(t)x}\phi_{i,k}^*(x)\hat{v}_{\mathrm{MF}}[\rho - \rho_0]\phi_{i,k}(x) e^{+\mathrm{i}A(t)x} \right].
 \label{eq:TD-qHF-energy}
\end{align} 
We can show that the equation of motion \eqref{eq:TD-qHF} is obtained as the functional derivative of the energy with respect to the orbital, as
\begin{align}
 \mathrm{i} \frac{\partial}{\partial t}\psi_{i,k}(x,t)
 = 
 \frac{1}{2}\frac{\delta E}{\delta \psi_{i,k}^*(x,t)}, \notag \\
 -\mathrm{i} \frac{\partial}{\partial t}\psi_{i,k}^*(x,t)
 = 
 \frac{1}{2}\frac{\delta E}{\delta \psi_{i,k}(x,t)},
\end{align}
where half of the prefactor in the energy functional derivative reflects that the total energy is for both spins and the orbital is for each spin.
These relations ensure that the energy \eqref{eq:TD-qHF-energy} is conserved in a time-dependent manner as long as the external field $A$ is a constant in time even for orbitals other than the initial one, because of 
\begin{align}
 \frac{\mathrm{d}}{\mathrm{d}t} E[\psi, \psi^*]
 = 
 \sum_{i,k}\int \! \mathrm{d}x \left[ \frac{\partial \psi_{i,k}^*(x,t)}{\partial t} \frac{\delta E}{\delta \psi_{i,k}^*(x,t)} \right. \notag \\
 \left. +
 \frac{\delta E}{\delta \psi_{i,k}(x,t)}\frac{\partial \psi_{i,k}(x,t)}{\partial t} \right] \notag \\
 +
 2\frac{\mathrm{d}A}{\mathrm{d}t}\sum_{i,k}\int \! \mathrm{d}x \psi_{i,k}^* (x,t) \left[ -\mathrm{i}\frac{\partial}{\partial x} + A(t) \right]\psi_{i,k}(x,t) \notag \\
 -
 \frac{1}{2} \iint \mathrm{d}x \mathrm{d}x' \frac{\partial \rho_0(x,x';t)}{\partial t} w(x',x) \rho(x',x;t) \notag \\
 =
 \frac{\mathrm{d}A}{\mathrm{d}t} \left\langle -\mathrm{i}\frac{\partial}{\partial x} + A(t) \right\rangle
 -
 \frac{1}{2} \left\langle \frac{\partial \rho_0(t)}{\partial t} w \right\rangle. 
\end{align}

Finally, we determine the orbital set that gives the energy minima for the energy functional \eqref{eq:TD-qHF-energy} without the field. The stationary condition of energy with respect to orthonormalized orbitals is given by a condition that $\delta\left(E - \sum_{nm,k'}\lambda_{nm,k}\left\langle \left. \tilde{\psi}_{n,k'}\right|\tilde{\psi}_{m,k'}\right\rangle\right)/\delta \tilde{\psi}_{i,k}(x)=0$, where $\lambda_{nm,k}$ is the Lagrange multiplier to ensure the orthogonality. Using the standard procedure to derive the Hartree-Fock equation, this condition leads to nonlinear eigenvalue equations, as 
\begin{align}
 \tilde{\epsilon}_{i,k}\tilde{\psi}_{i,k}(x)
 = 
 \left[-\frac{1}{2}\frac{\partial^2}{\partial x^2} + v(x) + \hat{v}_{\mathrm{MF}}[\tilde{\rho} - \rho_0]\right]\tilde{\psi}_{i,k}(x) \notag \\
 \tilde{\rho}(x,x')
 =
 2\sum_{i,k} \tilde{\psi}_{i,k}(x) \tilde{\psi}_{i,k}^*(x') \label{eq:qHF}.
\end{align}
In general, the eigenvector $\psi_{ik}$ should be determined by the self-consistent field manner, as in the original Hartree-Fock equation. Here, a trivial solution of Eq. \eqref{eq:qHF} is achieved by $\tilde{\psi}_{i,k} = \phi_{i,k}$, $\tilde{\rho} = \rho_0$ and $\tilde{\epsilon}_{i,k} = \epsilon_{i,k}$. Therefore, the initial condition of the TD-qHF equation $\phi_{i,k}$ gives a variationally stationary value of the energy functional \eqref{eq:TD-qHF-energy}.

\section{Simulation scheme}
\label{section:simulation_scheme}
We introduce an explicit discretization scheme for the spatial coordinate to solve Eq. \eqref{eq:TD-qHF}. LPP is expressed as the sum of plane waves that have reciprocal lattice momenta. Whole ingredients in the equation of motion are written as matrix elements evaluated as the reciprocal lattice momentum. A predictor-corrector trick in time propagation is needed to obtain a reasonable solution when the e-h interaction is switched on.

Then, we derive an explicit formula to determine the one-body potential. The band gap and the reduced effective mass are analytically determined by the potential amplitude and the lattice constant within the degenerated perturbation theory among two energy branches \cite{Kittel}. The time-dependent problem of the two-band assumption results in mapping to the Landau-Zener (LZ) type Hamiltonian when we neglect the e-h interaction.

We extract the tunneling rates from a real-time solution of TD-qHF. We solve the time-dependent equation of motion under a time-dependent field where the electric field is kept constant after a slow ramp-up field. The number of excited electrons (NEE) is evaluated by the explicit projection of the time-dependent wave function onto conduction bands. The tunneling rate is obtained as the slope of NEE as a function of time, which is validated as long as the slope is constant in time. Compared to a stationary assumption to obtain the rate, where different assumptions lead to different values \cite{Holthaus2000}, this tunneling rate evaluation is the most direct and concrete way to obtain the tunneling rate for a given system.

\subsection{Discretization for numerical simulation}
We here derive the exact expression of the LPP function in TD-qHF. We utilize plane wave expansion for LPP, as $u_{i,k}(x,t) = \sum_{G} e^{\mathrm{i}Gx}u_{i,k}(G,t), G=2\pi n/a, n = 0, \pm 1, \pm 2, \dots$. The equation of motion for LPP is given as 
\begin{align}
 \mathrm{i}\frac{\mathrm{d}}{\mathrm{d}t}u_{i,k}(G,t)
 =
 \frac{1}{2}\left([G + k + A(t) \right]^2 u_{i,k}(G,t) \notag \\
 + \sum_{G'} v_{G-G'} u_{i,k}(G',t)
 +
 \sum_{G'}v_{\mathrm{MF},G,G'}^{k}[\rho - \rho_0]u_{i,k}(G',t) ,
 \label{eq:EoM-LPP-TD-qHF} \\
 v_{\mathrm{MF},G,G'}^{k}[\varrho(t) ]
 =
 N_ka w^{q=0}(G-G')\eta(G-G',t),\notag \\
 -\frac{N_k a}{2} \sum_{q,H} w^{q}(H) \varrho^{k-q}(G-H, G'-H;t),
\end{align}
where relevant Fourier components are given as
\begin{align}
 w^k(G) = \frac{1}{N_k a} \int_0^{N_k a} \!\mathrm{d}x \ e^{-\mathrm{i}(k+G)x} w(x), \\
 v_G = \frac{1}{a} \int_0^a \!\mathrm{d}x \ e^{-\mathrm{i}Gx} v(x), \\
 \eta(G,t) = \frac{1}{a} \int_0^a \!\mathrm{d}x \ e^{-\mathrm{i}Gx} \varrho(x,x;t), \\
 \rho^{k}(G,G';t)
 =
 2 \sum_i u_{ik} (G,t) u_{ik}^*(G',t),\\
 \rho_0^{k}(G,G';t)
 \simeq
 2 \sum_i v_{i,k+A(t)} (G) v_{i,k+A(t)}^*(G').
\end{align}
The approximately equal sign in the last formula comes from the following approximation:
\begin{align}
 \frac{1}{N_k a}\int_0^{N_k a} \! \mathrm{d}x \ e^{\mathrm{i}\kappa x} = \frac{1}{\mathrm{i}N_k a \alpha}\left(e^{\mathrm{i}\kappa N_k a} - 1 \right) 
 \stackrel{N_k a \to \infty}{\longrightarrow} \delta_{\kappa,0}. \label{eq:approximationA}
\end{align}
The spatial integration \eqref{eq:approximationA} becomes exact when $\kappa$ is on a $k$-grid. Thus, the large $N_k$ limit ensures this approximation becomes accurate because we can find any $k+A(t)$ from the infinitely dense $k$-grids. The finite spatial integration for $w^k(G)$ leads to a spatial periodicity of the real space counterpart: $w_{N_k}(x) = \sum_{k,G} w^{k}(G) e^{\mathrm{i}(k+G)x}, w_{N_k}(x+N_k a) = w_{N_k}(x)$. The real space counterpart $w_{N_k}(x)$ becomes the original $w(x)$ when we take a large $N_k$ limit. The reciprocal grids for $G$ and $k$ are dense enough to obtain well-converged results. The shifted eigenvector $v_{i,k+A(t)} (G)$ is constructed by explicit diagonalization each time-step from the time-dependent Hamiltonian, $\tfrac{1}{2}[G+k+A(t)]^2\delta_{G,G'}+v_{G-G'}$.

\subsection{Time-evolution protocol}
We utilize a unitary matrix as the propagator of the orbital function, as 
\begin{align} 
 u_{i,k}(G, t +\Delta t)
 =
 \sum_{G'} U_{G,G'}(h^k) u_{i,k}( G',t), \\
 U_{G,G'}(h_k)
 =
 \left\langle G \left| \sum_{n} e^{-\mathrm{i}h_k \Delta t} \right| G' \right\rangle,
\end{align}
where $h^k$ is a $k$-dependent Hermitian matrix (similar to the Hamiltonian). The unitary matrix is obtained from an explicit diagonalization of $h^k$. The unitary nature guarantees norm conservation in the time-propagation regardless of $h^k$ choice.

We utilize a predictor-corrector (PC) scheme for the selection of $h^k$ \cite{Sato2015}:
\begin{align} 
 h^k 
 =
 \frac{1}{2}\left\{
 h_{\mathrm{qHF}}^k[\rho(t)](t)
 +
 h_{\mathrm{qHF}}^k[\rho^p](t+\Delta t) \right\}, \\
 h_{\mathrm{qHF}}^k[\varrho](t)
 =
 \frac{1}{2} 
 \left[ G + k + A(t) \right]^2\delta_{G,G'} \notag \\
 +
 v_{G-G'} + v_{\mathrm{MF},G,G'}^{k}[\varrho - \rho_0],
\end{align}
where $\rho_0$ is constructed with $A(t)$. $\rho^p$ is the predictor constructed by 
\begin{align} 
 u_{i,k}^p(G)
 =
 \sum_{G'} U_{G,G'}(h_{\mathrm{qHF}}^k[\rho(t)](t)) u_{i,k}( G',t), \notag \\
 \rho^p
 =
 \frac{2}{N_k}\sum_{i} 
 e^{+\mathrm{i}kx}u^p_{i,k}(x) u_{i,k}^{p*}(x')e^{-\mathrm{i}kx'} .
\end{align}
Thus, we need to perform unitary matrix construction and matrix-vector operation twice to move one step forward.

This PC scheme for the unitary matrix construction is mandatory for the qHF case, namely, the case with the electron-hole interaction. Results without PC lead to unphysical behavior even for a short time-step $\Delta t = 0.02 \mathrm{a.u.} = 0.484 \mathrm{as}$ when a large $\alpha$ is used in Eq. \eqref{eq:alpha_in_w}, although unitarity in the time-propagation is guaranteed by the construction. We confirm that $\Delta t = 0.2$ is usually small enough for our systems when we use the PC scheme.

\subsection{Band gap, reduced mass, and tunneling rate within Landau-Zener model for a monochromatic potential}
\label{section:gap_mass_LZmodel}
We concentrate on IES in this section to explain the reference dynamics. In principle, an expected electronic structure can be obtained by choosing the proper potential in the reference system. We give an explicit protocol to obtain an arbitrary band gap and reduced mass of the e-h pair from a one-body potential. The monochromatic spatial potential, having only one wave number, leads to the Hamiltonian of the Landau-Zener (LZ) model. The LZ model is utilized to provide our reference system with a tunneling rate estimation.

To keep our problem as simple as possible, we consider a system that has only one band gap in the BZ within the degenerated perturbation theory. Specifically, we choose fully occupied (unoccupied) states below (above) the gap and investigate the transition from the occupied valence band to the unoccupied conduction band. The single gap is achieved by the potential $ v(x) = v_0 \cos \left( 2 \pi n \frac{x}{a} \right), n = 1,2, \dots $. The Fourier transformation is given as
\begin{align}
 v_{G-G'} = v_0 \left( \delta_{G-G',n 2\pi /a} + \delta_{G-G',-n 2\pi /a}\right).
\end{align}
This potential connects three components ($G$, $G_{\pm}' = G \pm n 2\pi /a$) in Eq. \eqref{eq:EoM-LPP-TD-qHF} when the mean-field is neglected. The kinetic energy for the $G$ and $G_{\pm}'$ components degenerates at $k_{\pm} = -G \mp n \pi/a$. Another kinetic energy for the $G_{\mp}'$ component $\tfrac{9}{2}(n\pi/a)^2$ is far from the degenerated energy $\tfrac{1}{2}(n\pi/a)^2$. Hereafter, we denote $G'$ as either $G'_+$ or $G_-'$. Therefore, a $2\times2$ matrix representation for the time-dependent Shr\"{o}dinger equation is justified by
\begin{align}
 \mathrm{i}\frac{\mathrm{d}}{\mathrm{d} t}
 \begin{pmatrix}
 u_k(G,t) \\
 u_k(G',t)
 \end{pmatrix} \notag \\
 =
 \begin{pmatrix}
 \frac{1}{2}[G+k+A(t)]^2 & v_{G-G'} \\
 \left(v_{G-G'}\right)^* & \frac{1}{2}[G'+k+A(t)]^2
 \end{pmatrix}
 \begin{pmatrix}
 u_k(G,t) \\
 u_k(G',t)
 \end{pmatrix}. \label{eq:EoM-preLZ}
\end{align}
In the degenerate perturbation theory among two levels \cite{Kittel}, band gap $\Delta$ appears in accordance with the amount of $2|v_{G-G'_{\pm}}|$ at $k=-\frac{G+G'_{\pm}}{2}$.
To get rid of the $k$ squared term in the Hamiltonian, we introduce a time-dependent phase factor associated with the average energy $\frac{1}{4}\left\{[G+k+A(t)]^2 + [G'+k+A(t)]^2\right\}$ as $\left(\tilde{u}_\kappa^G(t) , \tilde{u}_\kappa^{G'}(t) \right)^T = e^{-\mathrm{i}\frac{1}{4}\left\{[G+k+A(t)]^2 +[G'+k+A(t)]^2\right\}} \times \left(u_\kappa^G(t) , u_\kappa^{G'}(t) \right)^T$, with a variable change $\kappa = k + \frac{G+G'}{2}$.

\begin{align}
 \mathrm{i}\frac{\mathrm{d}}{\mathrm{d} t}
 \begin{pmatrix}
 \tilde{u}_\kappa(G,t) \\
 \tilde{u}_\kappa(G',t)
 \end{pmatrix} \notag \\
 =
 \begin{pmatrix}
 \frac{1}{2}(G-G')[\kappa + A(t)] & v_{G-G'} \\
 \left(v_{G-G'}\right)^* & -\frac{1}{2}(G-G')[\kappa + A(t)]
 \end{pmatrix}
 \begin{pmatrix}
 \tilde{u}_\kappa(G,t) \\
 \tilde{u}_\kappa(G',t)
 \end{pmatrix}. \label{eq:EoM-LZ}
\end{align}
The shifted crystal momentum $\kappa$ means a crystal momentum measured from the degenerated point.

The DC electric field is obtained by $A(t) = -E_{\mathrm{DC}} t$.
By choosing a proper time origin, the crystal momentum $\kappa$ can be set to zero.
Thus, Eq. \eqref{eq:EoM-LZ} can be exactly mapped to the LZ model \cite{Vitanov}.
The adiabaticity parameter given in \onlinecite{Holthaus2000} is $\varepsilon = (G-G')E_{\mathrm{DC}}/(2\left| v^{G-G'}\right|^2)$.

Let us derive formulae for the effective mass \eqref{eq:EoM-LZ} written as $v_{G-G'}$ and $G-G'$. By diagonalizing the Hamiltonian in \eqref{eq:EoM-LZ} with $A=0$, we have two energy branches:
\begin{align}
 \epsilon_{\pm} (\kappa)
 =
 \pm \sqrt{\frac{1}{4}(G-G')^2\kappa^2 + \left| v_{G-G'}\right|^2}.
\end{align}
Regarding $+$ and $-$ as conduction and valence bands, the reduced effective mass $\mu$ reads
\begin{align} 
 \frac{1}{\mu}
 =
 \left(\frac{\mathrm{d}^2 \epsilon_+}{\mathrm{d}\kappa^2}\right)^{-1}_{\kappa=0}
 -
 \left(\frac{\mathrm{d}^2 \epsilon_-}{\mathrm{d}\kappa^2}\right)^{-1}_{\kappa=0}
 =
 \frac{(G-G')^2}{\Delta},
\end{align}
where $\Delta = 2\left| v_{G-G'}\right|$. The pure imaginary momentum that gives zero eigenvalues of the Hamiltonian in \eqref{eq:EoM-LZ} is $\kappa = \mathrm{i}\sqrt{\mu \Delta}$. The diabatic transition amount for $t=-\infty$ to $\infty$ is given as $e^{-\pi /\varepsilon}$ \cite{Holthaus2000}, with $1/\varepsilon =\sqrt{\mu \Delta}\Delta /(2 E_{\mathrm{DC}})$. This amount is the incremental value in the event that the vector potential sweep crosses $A(t) = 0$. This transition alternatively happens in crystals because the vector potential sweeps over Brillouin zones with the constant velocity $E_{\mathrm{DC}}$ (see Fig. \ref{fig:Schematic}). The period of the event is $2\pi/(aE_{\mathrm{DC}})$, as this is how long it takes to pass the first Brillouin zone once for the vector potential. Then, we obtain the tunneling rate as 
\begin{align}
 w_{\mathrm{LZ}} = \frac{aE_{\mathrm{DC}}}{2\pi}e^{-\pi \sqrt{\mu \Delta}\Delta /(2 E_{\mathrm{DC}}) }. \label{eq:wLZ}
\end{align}
The exponential dependence is the same as the semiclassical treatment of the Zener-tunneling \cite{Glutsch}. This LZ treatment does not include a resonance tunneling structure as a function of $E_{\mathrm{DC}}$ because there is no $1/E_{\mathrm{DC}}$ oscillation in \eqref{eq:wLZ}. To obtain a more qualitative rate, a finite time interval rather than $-\infty$ to $\infty$ should be taken into account for the LZ transition amount \cite{Holthaus2000} The finite interval $\tau_{\pm}$ in the normalized LZ formula in Eq. (61) in \onlinecite{Holthaus2000} for $|G-G'|= n \frac{2\pi}{a}$ is given as 
\begin{align}
 \tau_{\pm} = \pm \frac{1}{2n\mu} \label{eq:taupm_LZ},
\end{align}
where we use the normalized time $\tau/t = |G-G'|E_{\mathrm{DC}}/(2 \left|v_{G-G'}\right|) = E_{\mathrm{DC}}/(\sqrt{\mu \Delta}) $. Thus, a smaller $n$ and lighter $\mu$ leads to a better approximation of Eq. \eqref{eq:wLZ} to Zener tunneling.

While an interpretation of the Keldysh parameter combined with LZ treatment provides interesting insights, it does not relate to our problem directly. This consideration is given in more detail in Appendix \ref{section:appendixKeldysh}.

The tunneling formula \eqref{eq:wLZ} itself is for pairwise states and is not intended to rely on the spatial dimension of reciprocal space. In other words, the tunneling rate can be a fair value for systems that have different spatial dimensions. The Brillouin zone integration is required to obtain the actual amount of excited electrons for pairwise states per length, surface, or volume. We further need factor 2 because of spin-degeneracy.

We need to determine how many electrons are in the primitive cell. We assume a band insulator where the bottom $N_{\mathrm{occ}}$-bands are occupied for the initial condition in the reference system. The electron number per cell is $2N_{\mathrm{occ}}$ because of the spin-degeneracy. Our potential $v(x) = v_0 \cos (2\pi n x/a)$ gives a gap $2 v_0$ between the $n$-th and $n+1$-th bands according to the degenerated perturbation theory. Since we focus on Zener tunneling from valence top to conduction bottom, we set $n = N_{\mathrm{occ}}$.

\subsection{Tunneling rate evaluation from real-time calculation}
We apply a DC electric field after a smooth ramp-up to get rid of excitations due to the sudden field switch-on. The actual field shape is 
\begin{align}
 A(t) = 
 \begin{cases}
 -E_{\mathrm{DC}} T \left[ \frac{(t+T)^3}{T^3} - \frac{(t+T)^4}{2T^4}\right] (-T \leq t < 0)\\
 -E_{\mathrm{DC}} (t + T/2)(0\leq t)
 \end{cases},\label{eq:A-field}
\end{align}
as shown in the inset in Fig. \ref{fig:Schematic}. The electric field is obtained as the temporal derivative by $E(t) = -\dot{A}$. In the ramp-up region, the electric field exhibits a cubic function, $E_{\mathrm{DC}}\left[ 3 (t+T)^2/T^2 - 2 (t+T)^3/T^3 \right] (-T \leq t < 0)$, for connecting smoothly to the constant value. The field shape gives $\dot{E}(-T) = \dot{E}(0) =0$. We take $T=1000 = 24.19 \ \mathrm{fs} = \hbar/(0.171 \mathrm{eV})$ for the simulations over all parameters. The energetic dimension, 0.171 eV, is much less than the band gaps of the four investigated systems.

We define NEE as 
\begin{align}
 N_{\mathrm{ex}}(t)
 =
 \frac{2}{N_k}
 \sum_{i,b(\in \mathrm{unocc.}),k} \int_0^a \! \mathrm{d}x \ v_{b,k+A(t)}^*(x) u_{i,k}(x,t), \label{eq:NEE}
\end{align}
where the instantaneous eigenfunction $v_{b,k+A(t)}$ is obtained by an explicit diagonalization each time. For oscillating electric fields, projection onto the instantaneous eigenfunctions is more suitable for evaluating NEE under finite electric fields \cite{Otobe2008} than field-free eigenfunctions. NEE is normalized in a unit cell, i.e., it is equal to the total number of electrons in the cell when all electrons are excited from valence bands. To obtain the excited carrier density, we need to multiply the cell volume, area, or length to NEE.

We fit the slope value as the tunneling rate for the temporal evolution of NEE from a real-time solution of TD-qHF. The LZ transition happens once in a period $T_{\mathrm{B}} = 2\pi/(aE_{\mathrm{DC}})$, which is the period of the Bloch oscillation. The tunneling rate is not expected to be stationary for $0 \leq t < T_{\mathrm{B}}$ because the LZ transition occurs only once. The fitting is performed for the slope after $T_{\mathrm{B}}$. As shown in the figures in Sect. \ref{section:results}, most data show a nicely linear dependence on time for the regime.

\section{Results}
\label{section:results}
We prepare four systems imitating (a) a one-dimensional time-dependent Hartree-Fock solution (1DTDHF) \cite{Ikemachi2018}, (b) a boron-nitride sheet (BN-sheet), (c) $\alpha$-quartz, and (d) GaAs. The characteristics of the systems are (a) extreme conditions for electron-hole attraction strength, (b) a two-dimensional semiconductor, (c) a three-dimensional insulator, and (d) a three-dimensional semiconductor. Note that these simulations only utilize one-dimensional integration over the Brillouin zone. These different materials are imitated using only bandgap, effective mass, and exciton binding energy, as described in the following subsections. The effective Coulomb interaction is expected to capture the main role of e-h interaction for the tunneling dynamics. The remaining effect (e.g., dynamical symmetry) is missing within the one-dimensional treatment.

We change the reciprocal lattice size $2\pi /a$, the number of occupied bands $N_{\mathrm{occ}}$, and the off-diagonal component $v_{G-G'}$ to obtain the given bandgaps and effective masses of the reference systems. The number of occupied bands is set to $1$ for simplicity. We utilize a scaled electron-electron interaction, as 
\begin{align}
 w(x)
 =
 \frac{\alpha}{\sqrt{1+x^2}}. \label{eq:alpha_in_w}
\end{align}
The softened Coulomb potential has a long-range tail and no singularity at the origin.
The strength $\alpha$ is chosen such that the one-dimensional Wannier equation \eqref{eq:Wannier-RS} gives the correct exciton binding energy to reported values; see Appendix \ref{section:appendixB} for more detail.
Here, $E_{\mathrm{op}}$ is introduced as the optical gap from the Wannier equation. The exciton binding energy is obtained as $E_{\mathrm{op}} - \Delta$.
We utilized parameters $a$, $v$, and $\alpha$ (Table \ref{table:system_data}) and confirmed that they give $\mu, \Delta, E_{\mathrm{op}}$ in a large enough Brillouin zone sampling for static calculation.

\begin{table}[h!]
 \caption{Dataset and physical constants derived by the Wannier equation for the four systems. The number of occupied bands $N_{\mathrm{occ}}$ is always $1$. $a$, $v$, $\Delta$, and $E_{\mathrm{op}}$ are in the atomic unit. }
 \begin{tabular}{l|ccc|ccc|c}
 System & $a$ [a.u.] & $v$ [a.u.] & $\alpha$ & $\mu$ & $\Delta$ [a.u.] & $E_{\mathrm{op}} [a.u.]$ & Ref. \\
 \hline
 (a) 1DTDHF & 1.85 & 0.174 & 0.68 & 0.0303 & 0.3487 & 0.2105 & \onlinecite{Ikemachi2018}\\
 (b) BN-sheet & 7.05 & 0.143 & 0.21 & 0.360 & 0.2860 & 0.2131& \onlinecite{Ferreira2019}\\
 (c) $\alpha$-quartz & 6.4 & 0.167 & 0.059 & 0.347 & 0.3340 & 0.3219& \onlinecite{NekrashevichGritsenko}\\
 (d) GaAs & 5.0 & 0.028 & 0.01 & 0.0355 & 0.05600 & 0.05591& \onlinecite{Cohen1990} 
 \end{tabular}\label{table:system_data}
\end{table}

\subsection{1DTDHF}
The first system imitates the time-dependent Hartree-Fock results reported in \onlinecite{Ikemachi2018}.
This system is more or less artificial due to the one-dimensional space treatment with the TD-HF method, leading to a strong electron-hole attraction.
However, the strong effect provides insight into how the electron-hole interaction affects the dynamics.
The band gap and the reduced effective mass for quasiparticle spectra are $\Delta = 0.35 = 9.5 \ \mathrm{eV}$ and $\mu = 0.03$, as obtained by the self-consistent Hartree-Fock solution in \onlinecite{Ikemachi2018}.
The exciton binding energy is $3.8 \ \mathrm{eV}$ obtained through the linear-response spectra derived by the direct real-time solution of the TD-HF method.
The normalized time interval \eqref{eq:taupm_LZ} is given as $\tau_{\pm} = \pm 16.5$.

First, we fix the off-diagonal component $v = 0.174$ to reproduce the band gap. Then, the lattice constant is set to $a = 1.85$ to reproduce the reduced effective mass. Finally, the strength of the electron-electron interaction is determined as $\alpha = 0.68$ by a condition such that the Wannier equation \eqref{eq:Wannier-RS} gives the same binding energy as the TD-HF result in \cite{Ikemachi2018}.

We use 13 and 400 grids for real and reciprocal space, respectively, to obtain well-converged results. The size of the time-step is $0.02 = 0.484 \ \mathrm{as}$, which is quite a small value and only mandatory with e-h interaction.
The IES dynamics require $0.2 = 4.84 \ \mathrm{as}$ for the time-step.

\begin{figure}[h!]
\centering
\includegraphics[scale=0.45]{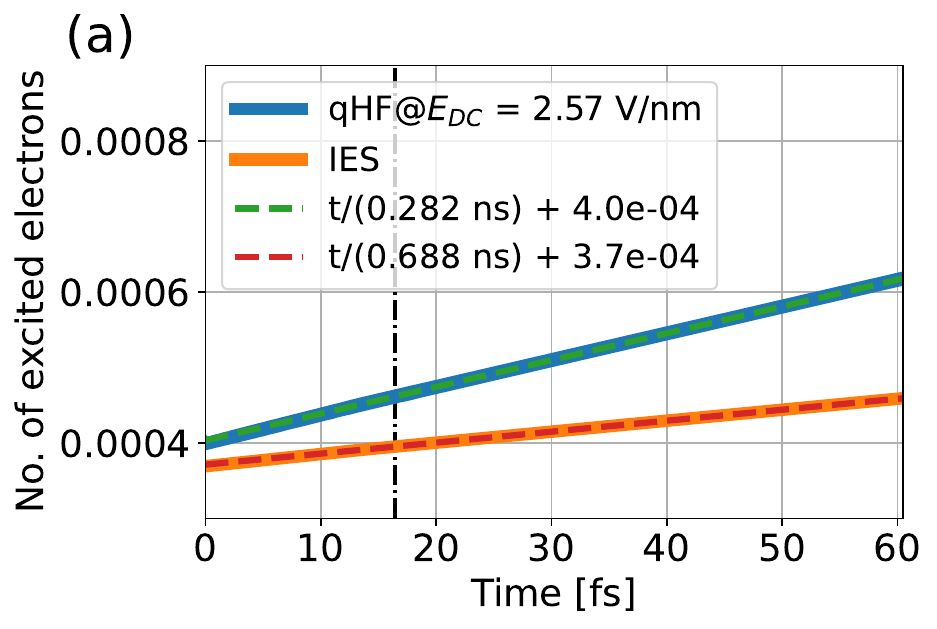}
\includegraphics[scale=0.45]{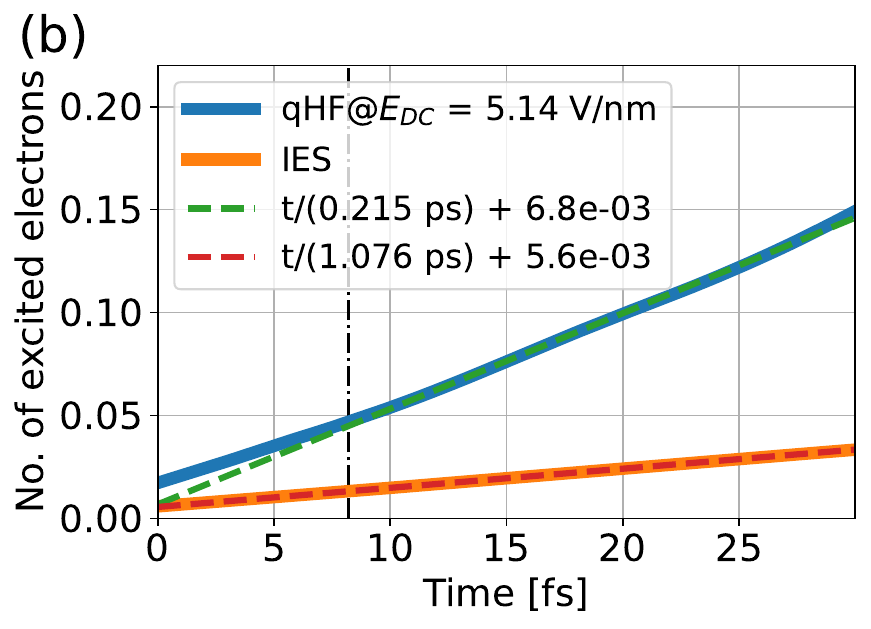}
\caption{Number of excited electrons as a function of time for 1DTDHF under (a) $E_{\mathrm{DC}} =2.6 \ \mathrm{V/nm}$ and (b) $E_{\mathrm{DC}} =5.1 \ \mathrm{V/nm}$. $T_{\mathrm{B}}$, 16.52 fs, and 8.26 fs are represented as vertical lines.}
\label{fig:1DTDHF}
\end{figure}

The time evolution of NEE for the DC field is presented in Fig. \ref{fig:1DTDHF}, where both graphs show linear behavior as a function of time after $T_{\mathrm{B}}$.
There are finite y-intercepts because the ramp-up field influences the dynamics for $t<0$.
By linear fitting with $w t + n$ for $0.005 = 2.57 \ \mathrm{V/nm}$ excitation, the tunneling rates $w$ for TD-qHF and IES are $8.58 \times 10^{-8} = 1/(0.282 \ \mathrm{ns})$ and $3.52 \times 10^{-8} = 1/(0.688 \ \mathrm{ns})$.
TD-qHF shows a 2.44 times more significant tunneling rate than IES. For $0.01 = 5.14 \ \mathrm{V/nm}$ excitation, the tunneling rates are $1.12 \times 10^{-4} = 1/(0.215 \ \mathrm{ps})$ and $2.25 \times 10^{-5} = 1/( 1.076 \ \mathrm{ps})$, and thus the enhancement factor is 4.98.
The tunneling rates and the field-dependent enhancements are summarized in Table \ref{table:1DTDHF}. The LZ formula \eqref{eq:wLZ} gives a quantitatively accurate estimation for IES, which can be attributed to the light-reduced mass.

\begin{table}[h!] 
 \caption{Calculated tunneling rates and enhancement factors for 1DTDHF system. The period of Bloch oscillation and the tunneling rates evaluated with LZ formula \eqref{eq:wLZ} are tabulated.}
 \begin{tabular}{ccc|ccc}
 $E_{\mathrm{DC}}$ [a.u.]& $T_{\mathrm{B}}[a.u.]$ & 2$w_{\mathrm{LZ}}$ [a.u.]& IES [a.u.]& qHF [a.u.]& Enhancement \\
 \hline
 0.003 & 1130 & 1.25e-11 & 1.16e-11 & 2.29e-11 & 1.97 \\
 0.005 & 680 & 3.79e-8 & 3.52e-8 & 8.58e-8 & 2.44 \\
 0.01 & 340 & 2.11e-5 & 2.25e-5 & 1.12e-4 & 4.98
 \end{tabular}
 \label{table:1DTDHF}
\end{table}

NEE is a normalized number in a cell.
When the field is 2.57 V/nm, NEE reaches 0.1\% of the total electron after 1 ps, where two electrons are in the cell.
After doubling the strength to 5.14 V/nm, NEE reaches 0.1\% after 1 fs.
The abrupt increase of NEE with respect to the field increase stems from the strong nonlinearity in the tunneling process.

\subsection{BN-sheet}
The second system imitates a BN-sheet, which is a two-dimensional semiconductor.
The protocol to determine $v$, $a$, and $\alpha$ as the material parameters is the same as for 1DTDHF.
According to \onlinecite{Ferreira2019}, $\Delta = 0.2860 = 7.77 \ \mathrm{eV}$ and $\mu = 0.36$ derived by effective masses for hole $m_h = 0.63$ and particle $m_e = 0.83$.
The binding energy of excitons is 2.14 eV. Then, we utilize the following parameters for the BN-sheet: $v = 0.143$, $a = 7.05$, and $\alpha = 0.21$.
The normalized time interval \eqref{eq:taupm_LZ} is given as $\tau_{\pm} = \pm 1.39$.

We use 53 and 140 grids for real and reciprocal space, respectively, to obtain well-converged results. The time-step size is 0.2, the same as the IES of 1DTDHF.

\begin{figure}[h!]
 \centering
 \includegraphics[scale=0.45]{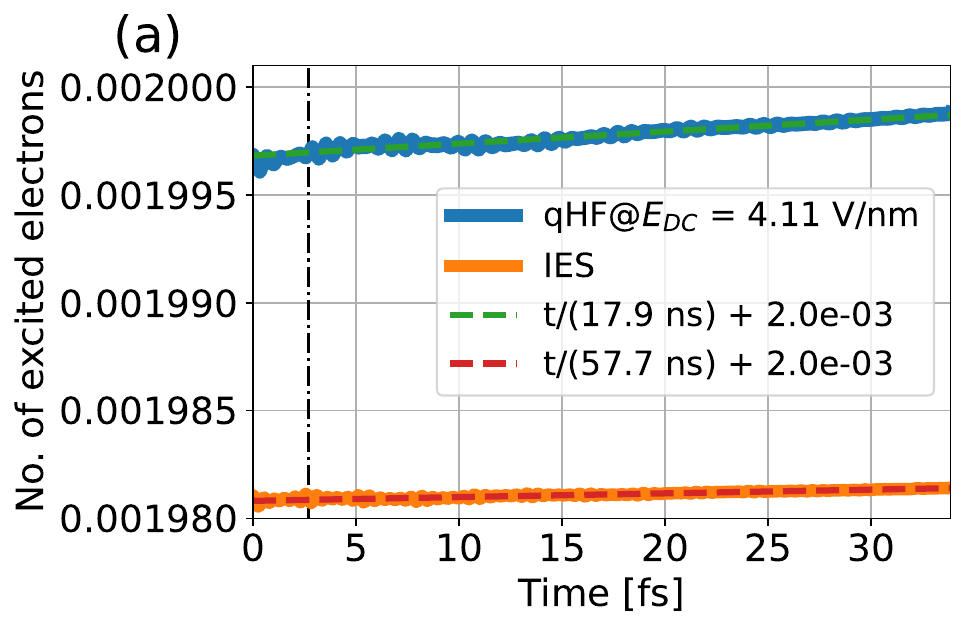}
 \includegraphics[scale=0.45]{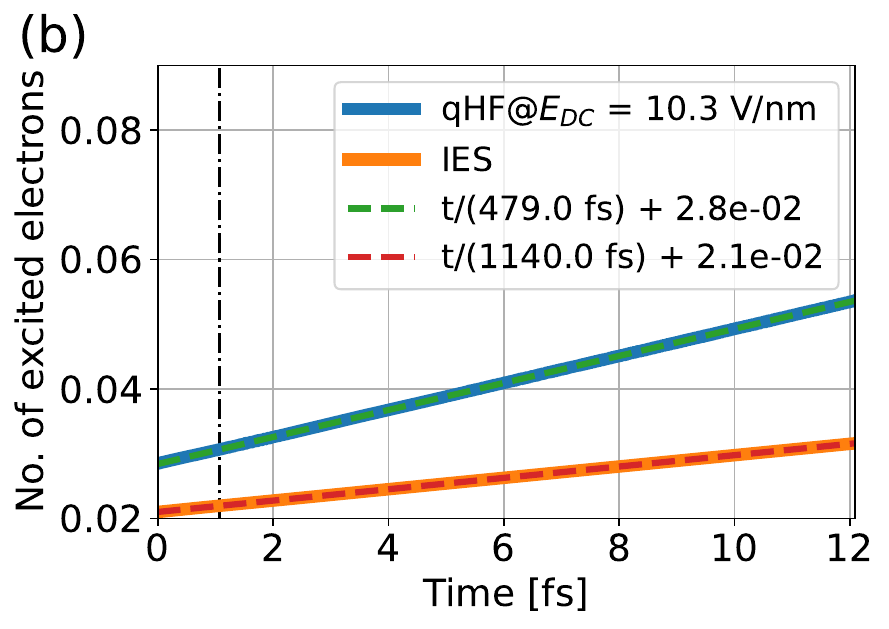}
 \caption{Number of excited electrons as a function of time for BN-sheet under (a) $E_{\mathrm{DC}} =4.1 \ \mathrm{V/nm}$ and (b) $E_{\mathrm{DC}} =10.3 \ \mathrm{V/nm}$. $T_{\mathrm{B}}$, 2.70 fs, and 1.078 fs are represented as vertical lines.}
 \label{fig:BN}
\end{figure}

Figure \ref{fig:BN} shows the time-dependent NEE values for the BN-sheet system. The tunneling rates are obtained by the same procedure as the 1DTDHF system.
For 4.1 V/nm excitation, the tunneling rates are $1.35 \times 10^{-10} = 1/(17.9 \ \mathrm{ns})$ and $4.19 \times 10^{-9} = 1/(57.7 \ \mathrm{ns})$, and thus the enhancement factor is 3.22.
Both tunneling rates are $5.05 \times 10^{-5} = 1/(479 \ \mathrm{fs})$ and $2.12 \times 10^{-5} = 1/(1140 \ \mathrm{fs})$ for 10.3 V/nm excitation.
Table \ref{table:BN} summarizes the tunneling rates and field-dependent enhancements.
The ionization rate of qHF with the $E_{\mathrm{DC}} = 0.01 $ field has significant ambiguity because the NEE does not show a strict linear dependence on time.

\begin{table}[h!]
 \caption{Calculated tunneling rates and enhancement factors for BN-sheet system. The period of Bloch oscillation and the tunneling rates evaluated with LZ formula \eqref{eq:wLZ} are tabulated. (?) symbols in the qHF and enhancement rows with $E_{\mathrm{DC}}=0.01$ mean the corresponding NEE does not show a strict linear line.}
 \begin{tabular}{ccc|ccc}
 $E_{\mathrm{DC}}$ [a.u.]& $T_{\mathrm{B}}[a.u.]$ & 2$w_{\mathrm{LZ}}$ [a.u.]& IES [a.u.]& qHF [a.u.]& Enhancement \\
 \hline
 0.008 & 111 & 2.68e-10 & 4.19e-10 & 1.35e-9 & 3.22\\
 0.009 & 99.0 & 2.23e-9 & 1.47e-9 & 2.84e-9 & 1.93 \\
 0.01 & 89.1 & 1.23e-8 & 2.84e-7 & 1.41e-7(?) & 0.496(?) \\
 0.015 & 59.4 & 2.26e-6 & 2.16e-6 & 6.44e-6 & 2.98 \\
 0.02 & 44.6 & 3.33e-5 & 2.12e-5 & 5.05e-5 & 2.38
 \end{tabular}
 \label{table:BN}
\end{table}

Compared to the 1DTDHF system, the BN-system is barely excited for both IES and qHF, while the band gap is smaller.
This is due to the much heavier reduced mass than the 1DTDHF system.
Because of this heavier mass, the LZ formula occasionally fails to predict the ionization rate, in contrast to the 1DTDHF system.
When we assume the field comes from an optical field, the field intensity for 10.3 V/nm is 14 $\mathrm{TW/cm^2}$.
According to a rough estimation that just a half-period of a single oscillation effectively ionizes the electron into the conduction band by the tunneling rate, 1 \% of photocarrier is generated in 7.5 fs with 14 $\mathrm{TW/cm^2}$ intensity.

\subsection{$\alpha$-quartz}
The third system imitates $\alpha$-quartz, a three-dimensional insulator.
The protocol to determine $v$, $a$, and $\alpha$ as the material parameters is the same as before.
According to \onlinecite{NekrashevichGritsenko}, $\Delta = 0.334 = 9.1 \ \mathrm{eV}$ and $\mu = 0.347$ derived by effective masses for hole and particle are $m_h = 1.3$ and $m_e = 0.5$.
The binding energy of excitons is estimated to be 0.33 eV, since the Hydrogen 1S state energy with the screened Coulombic interaction with dielectric constant $\epsilon_r \sim 3.8$ for the reduced mass by $ E_n = - \frac{\mu}{2\epsilon_r ^2 }$. Then, we utilize the following parameters for $\alpha$-quartz: $v = 0.167$, $a = 6.4$, and $\alpha = 0.059$.
The normalized time interval \eqref{eq:taupm_LZ} is given as $\tau_{\pm} = \pm 1.44$.

We use 103 and 40 grids for real and reciprocal space, respectively, to obtain well-converged results. The time step size is 0.2 a.u., the same as the BN-sheet.

\begin{figure}[h!]
 \centering
 \includegraphics[scale=0.45]{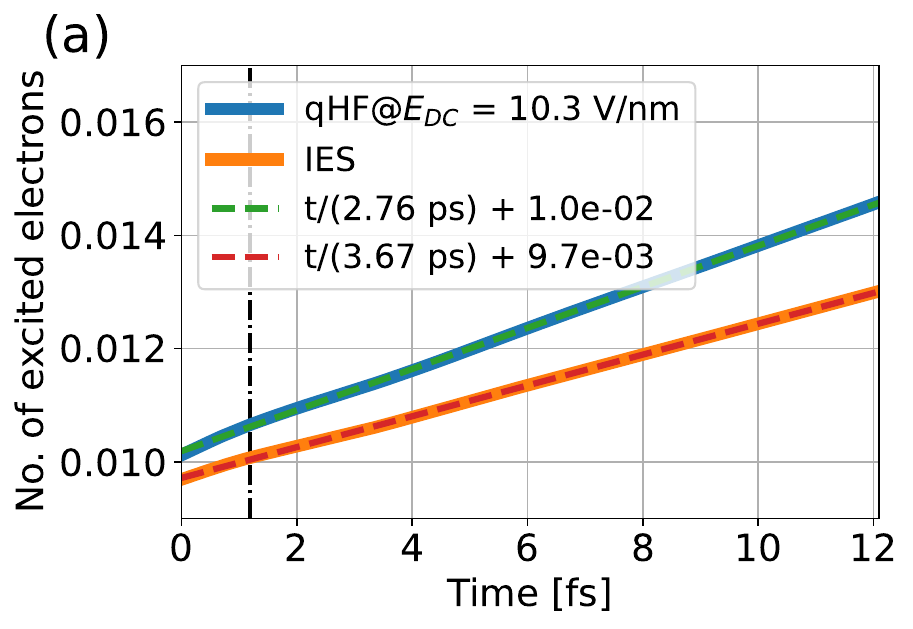}
 \includegraphics[scale=0.45]{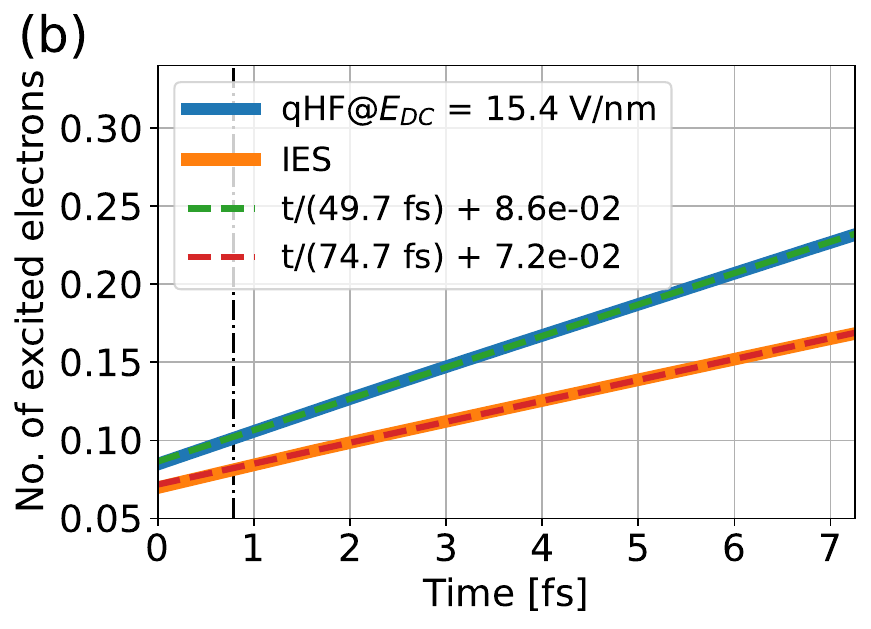}
 \caption{Number of excited electrons as a function of time for $\alpha$-quartz under (a) $E_{\mathrm{DC}} =10.3 \ \mathrm{V/nm}$ and (b) $E_{\mathrm{DC}} =15.4 \ \mathrm{V/nm}$. $T_{\mathrm{B}}$, 1.187 fs, and 0.792 fs are represented as vertical lines.}
 \label{fig:SiO2}
\end{figure}

Figure \ref{fig:SiO2} shows time-dependent NEE values for the $\alpha$-quartz system.
We used the same procedure for obtaining the tunneling rates as the 1DTDHF system.
For $0.02 = 10.3 \ \mathrm{V/nm}$ excitation, the tunneling rates are $8.77 \times 10^{-6} = 1/(2.76 \ \mathrm{ps})$ and $6.58 \times 10^{-6} = 1/(3.67 \ \mathrm{ps})$, and thus the enhancement factor is 1.33.
Both tunneling rates are $4.86 \times 10^{-4} = 1/(49.7 \ \mathrm{fs})$ and $3.24 \times 10^{-4} = 1/(74.7 \ \mathrm{fs})$ for $0.03 = 15.4 \ \mathrm{V/nm}$ excitation.
Table \ref{table:SiO2} summarizes the tunneling rates and field-dependent enhancements.

\begin{table}[h!]
 \caption{Calculated tunneling rates and $\alpha$-quartz system enhancement factors. The period of Bloch oscillation and the tunneling rates evaluated with LZ formula \eqref{eq:wLZ} are tabulated.}
 \begin{tabular}{ccc|ccc}
 $E_{\mathrm{DC}}$ [a.u.]& $T_{\mathrm{B}}[a.u.]$ & 2$w_{\mathrm{LZ}}$ [a.u.]& IES [a.u.]& qHF [a.u.]& Enhancement \\
 \hline
 0.01 & 98.2 & 3.57e-10 & 2.59e-10 & 3.00e-10 & 1.16 \\
 0.02 & 49.1 & 5.39e-6 & 6.58e-6 & 8.77e-6 & 1.33\\
 0.03 & 32.7 & 1.59e-4 & 3.24e-4 & 4.86e-4 & 1.50
 \end{tabular}
 \label{table:SiO2}
\end{table}

The tendency of the $\alpha$-quartz system is similar to that of the BN-sheet system, with a relatively heavy mass and relatively high band gap.
One of the most significant differences is the much smaller e-h interaction, mainly due to the high dimensionality.
The enhancement of e-h is in the range from 16 to 50 \% increase in the investigated strength.

\subsection{GaAs}
The final system imitates GaAs, a three-dimensional semiconductor.
The protocol to determine $v$, $a$, and $\alpha$ as the material parameters is the same.
According to \onlinecite{Cohen1990}, $\Delta = 1.519 \ \mathrm{eV}$ and $\mu = 0.0377$ derived by the effective masses for light-hole and particle are $m_h = 0.087$ and $m_e = 0.0665$.
The binding energy of excitons is estimated to be 0.003 eV since the 1S state energy with the dielectric constant $\sim 13$ for the reduced mass.
Then, we utilize the following parameters for GaAs: $v = 0.028$, $a = 5.0$, and $\alpha = 0.01$. The normalized time interval \eqref{eq:taupm_LZ} is given as $\tau_{\pm} = \pm 14.1$.

We use 13 and 140 grids for real and reciprocal space, respectively, to obtain well-converged results. The time-step size is 0.2 a.u., the same as the BN-sheet and $\alpha$-quartz.

\begin{figure}[h!]
 \centering
 \includegraphics[scale=0.45]{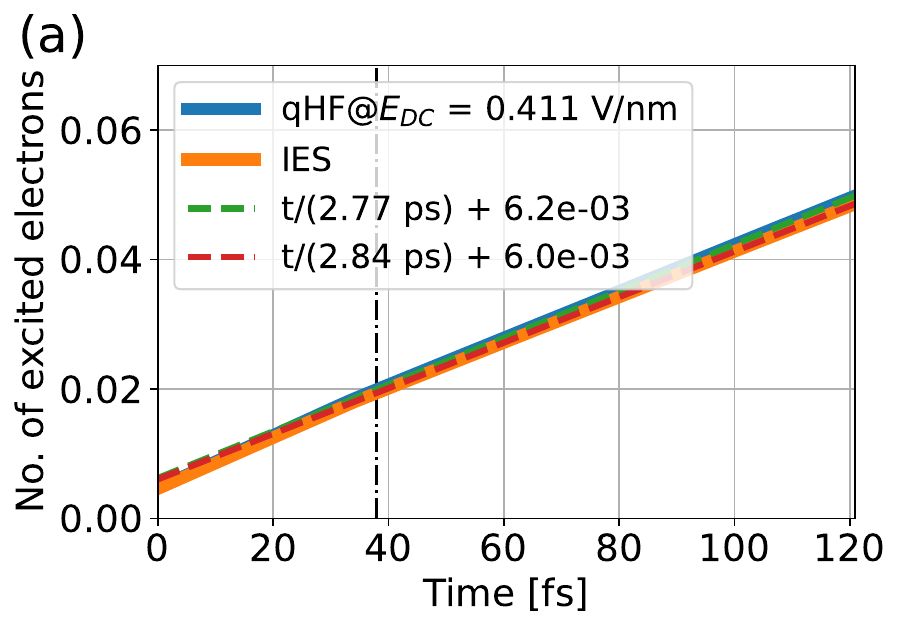}
 \includegraphics[scale=0.45]{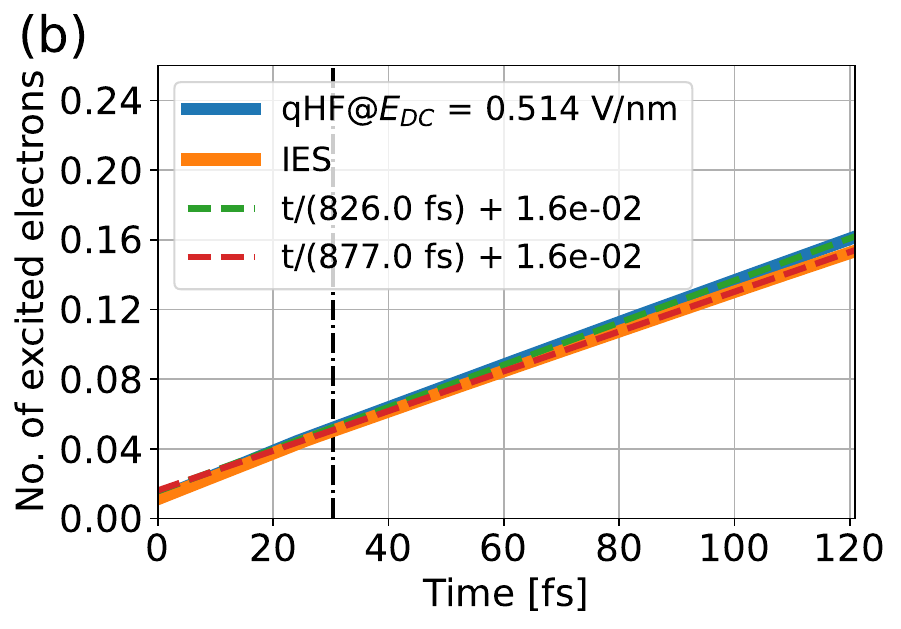}
 \caption{Number of excited electrons as a function of time for GaAs under (a) $E_{\mathrm{DC}} =0.411 \ \mathrm{V/nm}$ and (b) $E_{\mathrm{DC}} =0.514 \ \mathrm{V/nm}$. $T_{\mathrm{B}}$, 38.0 fs, and 30.4 fs are represented as vertical lines.}
 \label{fig:GaAs}
\end{figure}

Figure \ref{fig:GaAs} shows the time-dependent NEE values for the GaAs system.
We used the same procedure to obtain the tunneling rates as the 1DTDHF system. For $0.0008 = 0.411 \ \mathrm{V/nm}$ excitation, the tunneling rates are $8.73 \times 10^{-6} = 1/(2.77 \ \mathrm{ps})$ and $8.52 \times 10^{-6} = 1/(2.84 \ \mathrm{ps})$, and thus the enhancement factor is 1.02.
Both tunneling rates are $2.93 \times 10^{-5} = 1/(826 \ \mathrm{fs})$ and $2.76 \times 10^{-5} = 1/(877 \ \mathrm{fs})$ for $0.001 = 0.514 \ \mathrm{V/nm}$ excitation.
Table \ref{table:GaAs} summarizes the tunneling rates and field-dependent enhancements.
The LZ formula \eqref{eq:wLZ} gives a good estimation for IES because of the light-reduced mass of GaAs.

\begin{table}[h!]
 \caption{Calculated tunneling rates and enhancement factors for GaAs system. The period of Bloch oscillation and the tunneling rates evaluated with LZ formula \eqref{eq:wLZ} are tabulated.}
 \begin{tabular}{ccc|ccc}
 $E_{\mathrm{DC}}$ [a.u.]& $T_{\mathrm{B}}[a.u.]$ & 2$w_{\mathrm{LZ}}$ [a.u.]& IES [a.u.]& qHF [a.u.]& Enhancement \\
 \hline
 0.0005 & 2510 & 3.12e-7 & 3.62e-7 & 3.64e-7 & 1.01\\
 0.0008 & 1570 & 9.46e-6 & 8.52e-6 & 8.73e-6 & 1.02\\
 0.001 & 1260 & 3.15e-5 & 2.76e-5 & 2.93e-5 & 1.06 \\
 \end{tabular}
 \label{table:GaAs}
\end{table}

The NEE of the GaAs system reached a few electrons in a cell with a much weaker field than the other three systems. This is because GaAs has a light-reduced mass and a much smaller gap. Similar tunneling rates are achieved with a field strength tens of times weaker than the $\alpha$-quartz system. Reflecting the weaker e-h interaction strength, the enhancement due to e-h interaction is minor, up to 0.514 V/nm.

\section{Discussion}
\label{section:discussion}
\begin{figure}[h!]
 \centering
 \includegraphics[scale=0.45]{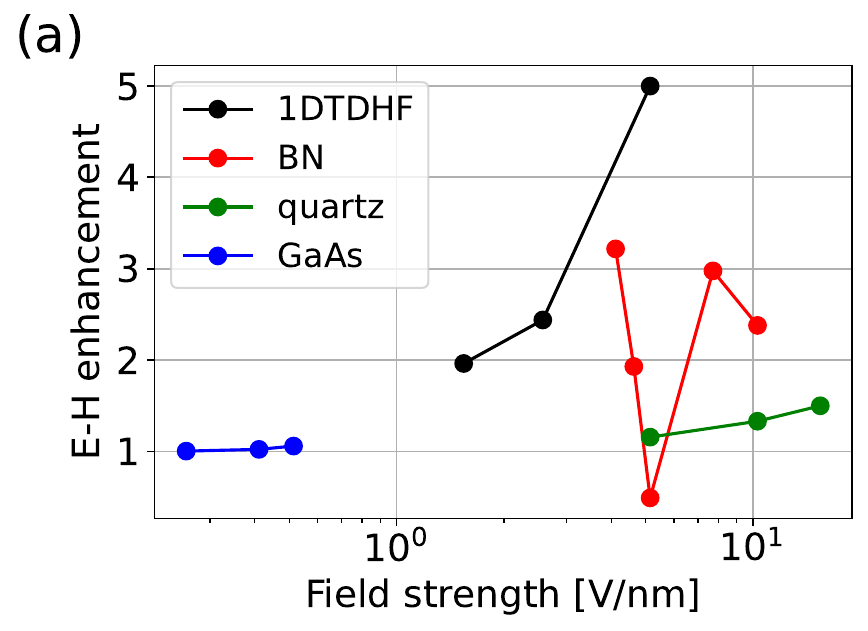}
 \includegraphics[scale=0.45]{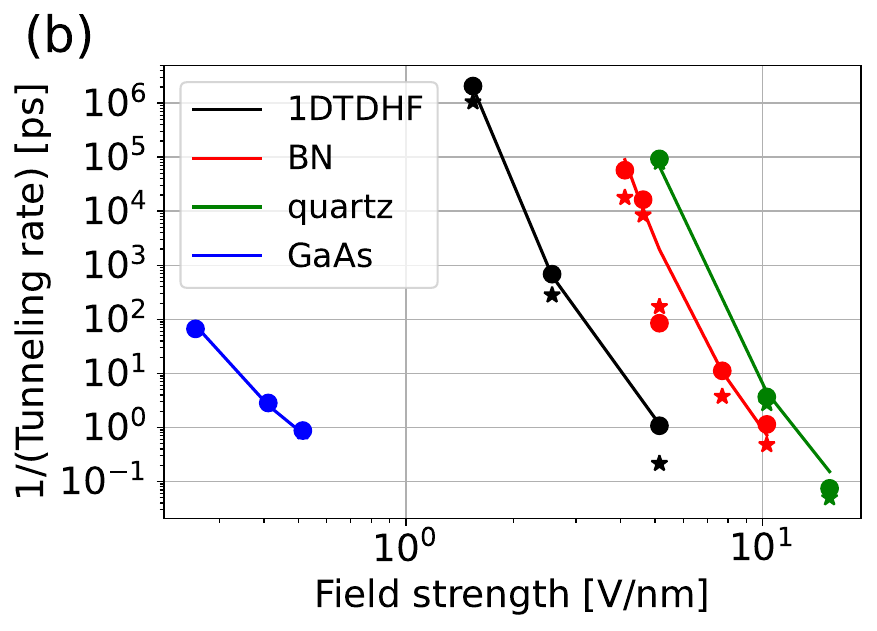}
 \caption{(a) Enhancement factors and (b) tunneling rate inverse as a function of field strength for 1DTDHF (black), BN-sheet (red), $\alpha$-quartz (green), and GaAs (blue). LZ-based formula, IES, and qHF data are denoted by solid lines, solid circles $\bullet$, and solid stars $\star$, respectively. }
 \label{fig:Summary}
\end{figure}

We examine the trend of the enhancement due to e-h interaction and the tunneling rate over the investigated materials and field, as shown in Fig. \ref{fig:Summary}.
The enhancement factors in Fig. \ref{fig:Summary}(a) are normally larger than unity except for the BN-sheet system at $0.01 = 5.14 \mathrm{V/nm}$, at which ionization rates are not rigorously defined because NEE does not show a clear straight line.
We find a tendency that significant enhancements are accompanied by a large e-h  interaction strength $\alpha$.
The enhancement factors increase with field strength increasing for the 1DTDHF, $\alpha$-quartz, and GaAs systems.

The LZ formula basically gives us a reasonable estimation of the tunneling rates for all examples in Fig. \ref{fig:Summary}(b). Specifically, it shows quite a good estimation for IES when the reduced mass is light, as in the 1DTDHF, and GaAs systems. For the BN-sheet system, the LZ formula maximally predicts an order error for a heavier mass of around 0.35. The e-h enhancement increases for a stronger field with the 1DTDHF, $\alpha$-quartz, and GaAs systems.

Here, we discuss the mechanism underlying the enhancement due to e-h interaction.
Band-gap renormalization (BGR) can be explained as occurring through carrier introduction by photodoping, thermal excitation, or chemical doping.
These gap reductions can be explained as the exchange interaction weakening by Eq. \eqref{eq:gap-renormalization_SBE} in two-band SBE.
Thus, the possible mechanism for enhancing tunneling rates is attributed to electronic structure renormalization.
To confirm this mechanism, we perform on-the-fly electronic structure extraction from time-dependent qHF Hamiltonian $h^k[\rho(t)](t)$.
A system that has light effective mass is suitable to evaluate the tunneling rate using the LZ formula.
We analyze the data for the 1DTDHF system because of its light mass and strong e-h interaction strength.

\begin{figure}[h!]
 \centering
 \includegraphics[scale=0.45]{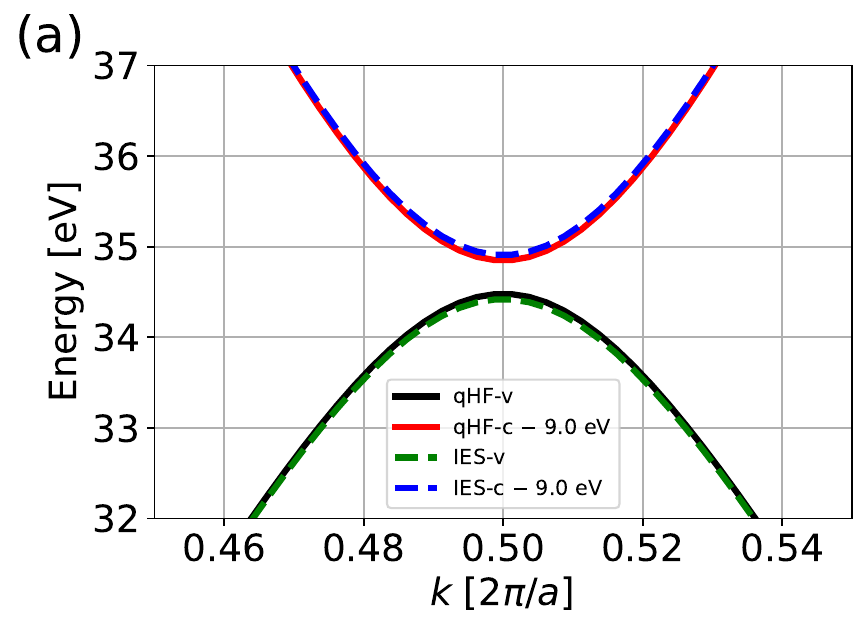}
 \includegraphics[scale=0.45]{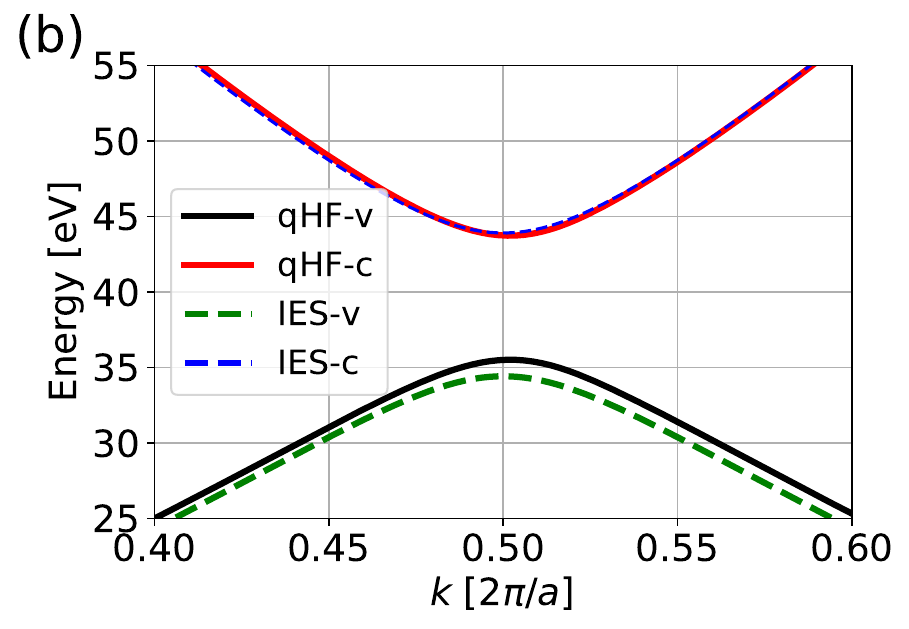}
 \caption{Snapshots of transient electronic structures for 1DTDHF system under 5.14 V/nm at $t= 60 \ \mathrm{fs}$ (a) and 10.3 V/nm at $t= 25 \ \mathrm{fs}$ (b). The conduction band energy in the top panel is shifted downward by 9.0 eV for visibility.}
 \label{fig:1DTDHF_electronic_structure}
\end{figure}

We evaluate a time-dependent eigenvalue $\mathcal{E}_i^k(t)$ by the explicit diagonalization of $h^k[\rho(t)](t)$, where $i$ refers to the eigenvalue index for each $k$.
Note that this eigenvalue is not precisely equal to the diagonal component of the generalized Rabi frequency in Eq. \eqref{eq:Generalized_Rabi_frequency_TD-qHF} because each use a different basis to evaluate the value.
Snapshots of the time-dependent eigenvalue for the 1DTDHF system are shown in Fig. \ref{fig:1DTDHF_electronic_structure}.
For the reference system, IES, the eigenvalue shows an $A$-shifted relation $\mathcal{E}_i^k(t)=\epsilon_{i,k+A(t)}$ stemming from the velocity gauge coupling.
The trivial momentum shift is subtracted to make the data in the first Brillouin zone.
For the weaker field case, 5.14 V/nm, the eigenvalue change appears symmetric for the valence and conduction bands.
We also extract the reduced mass from $\mathcal{E}_i^k(t)$ by $1/\mu(t) = \left(\partial^2 \mathcal{E}_c^k(t)/\partial k^2\right)^{-1}_{k=\pi/a} - \left(\partial^2 \mathcal{E}_v^k(t)/\partial k^2\right)^{-1}_{k=\pi/a}$.
The amount of the renormalized gap and reduced mass are 9.375 eV and 0.0295, respectively.
The renormalized electronic structure shows the asymmetric change for the stronger field case, 10.3 V/nm: valence band uplift is more pronounced than the conduction band drop. The amount of the renormalized gap and reduced mass are 8.3 eV and 0.027.
Combining the renormalized gap and effective masses into the LZ-base tunneling formula \eqref{eq:wLZ}, we obtain 1.2 and 3.7 enhancements for 5.14 V/nm and 10.3 V/nm.
There needs to be more than this estimation for the actual ionization rate enhancement, 2.44 and 4.98 in Table \ref{table:1DTDHF}.
This fact suggests that renormalization of the transient dipole moment also plays a role in the enhancement.
The hauling-up effect is a possible candidate for the dipole moment renormalization.

\begin{figure}[h!]
 \centering
 \includegraphics[scale=0.45]{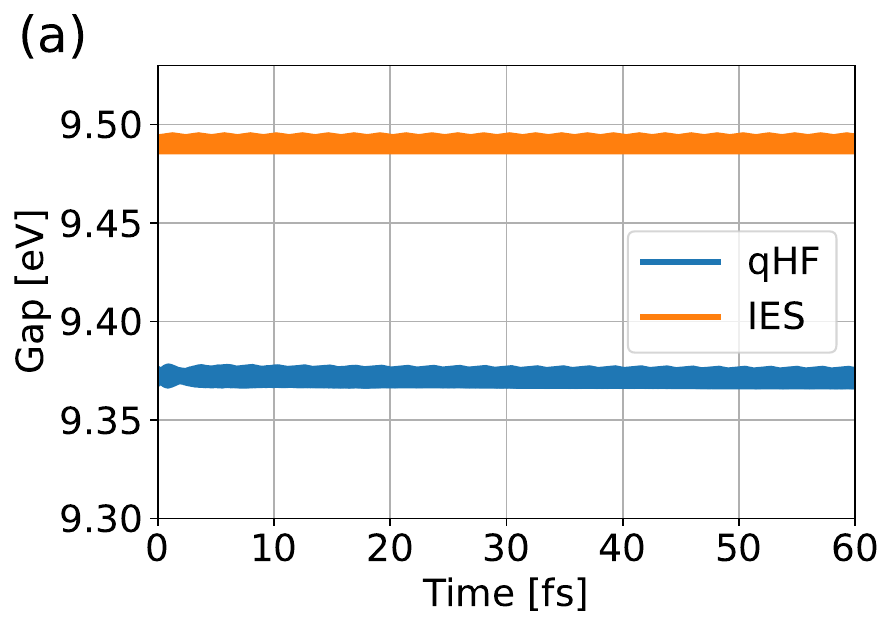}
 \includegraphics[scale=0.45]{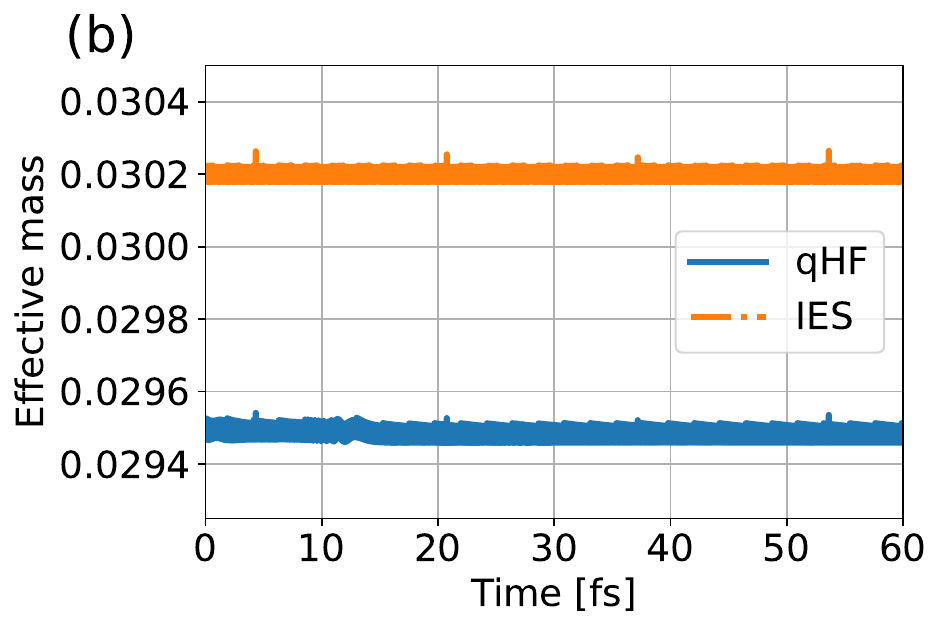}
 \includegraphics[scale=0.45]{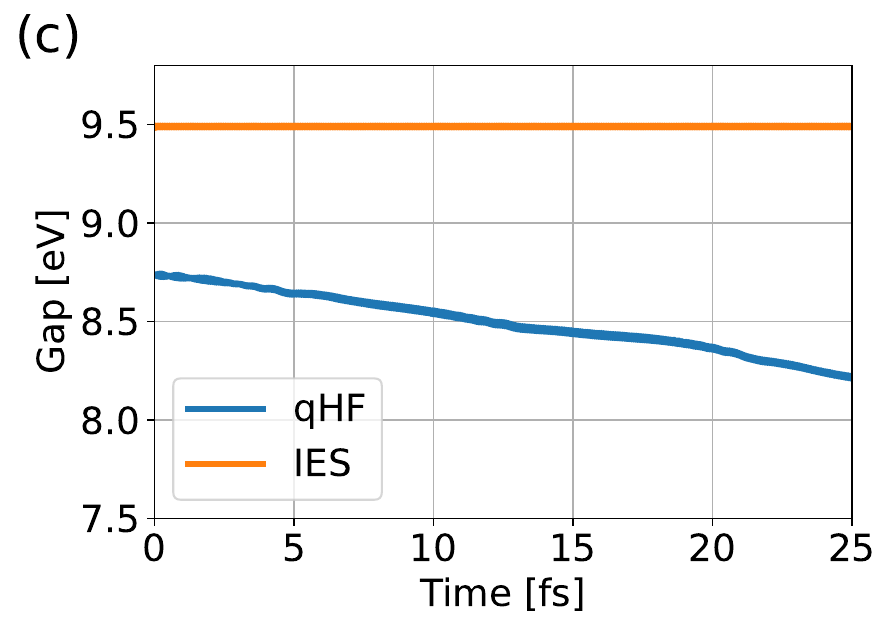}
 \includegraphics[scale=0.45]{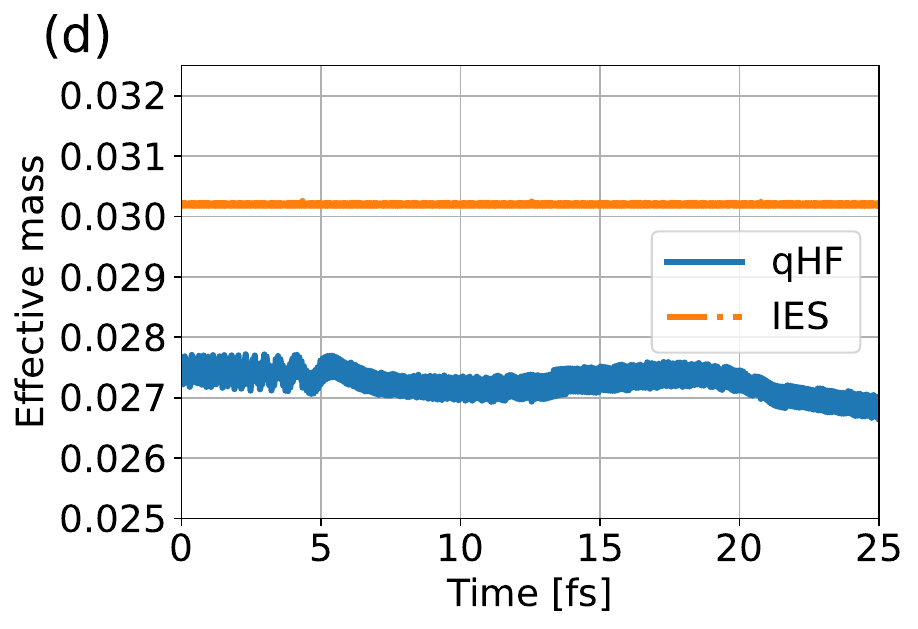}
 \caption{Time evolution of transient gaps (left) and reduced masses (right) for 5.14 V/nm (top) and 10.3 V/nm (bottom).}
 \label{fig:1DTDHF_electronic_structure_evolution}
\end{figure}

The time-dependent renormalized gap and reduced mass are also instructive for understanding the dynamics, as shown in Fig. \ref{fig:1DTDHF_electronic_structure_evolution}. For the weaker field case, neither the gap nor the reduced mass change over time. In contrast, the renormalized gap gradually decreases as a function of time for the stronger field case. The renormalized reduced mass is independent of time. These facts demonstrate that the gap reduction depends on instantaneous field strength and the history of the time-dependent electric field.

\section{Conclusion}
\label{section:conclusion}
We developed a TD-qHF theoretical framework to investigate the role of e-h interaction in solid-state electron dynamics driven by a time-dependent electric field. The e-h interaction is included as a mean field inspired by the TD-HF method. By subtracting the trivial ground state DM in the mean-field potential, an arbitrary IES can be introduced as a reference system. The strength of the e-h interaction can be scaled up or down without any influence on IES. This independent control of the IES and e-h interaction strength cannot be achieved by the original TD-HF method. As such, this flexible property of TD-qHF allows us to model a system more easily than TD-HF. The equation of motion can be derived via the variational principle from an explicit energy expression; the total energy is well-defined when an external field is absent.

We performed TD-qHF simulations to evaluate the e-h attraction effect in Zener tunneling. Thanks to explicit time evolution, there was no ambiguity in handling the nonperturbative response of quantum systems. A mapping to the LZ model of IES time evolution was derived for the tunneling rate estimation. The influence of e-h interaction was investigated by comparing the TD-qHF and IES simulations using four systems imitating 1DTDHF, BN-sheet, $\alpha$-quartz, and GaAs.

Our findings showed that e-h interaction enhances the ionization rate in almost all cases. The enhancement depends on the systems as well as on the field strength. For the 1DTDHF, $\alpha$-quartz, and GaAs systems, the enhancement showed a monotonic increase as a function of the field strength. We analyzed the transient band structure change due to e-h interaction for the 1DTDHF system and found that the interaction reduces the gap and the reduced mass, leading to tunneling rate promotion. However, the reductions of the gap and mass do not fully explain the enhancement. The remaining dynamical effects, such as the hauling-up effect, play an additional role in the enhancement. This theoretical framework and the results will help to clarify the mechanisms and control the dynamics of two-dimensional semiconductors (e.g., transition metal dichalcogenides) driven by a strong light field.

\begin{acknowledgments}
Y. S. thanks Kohei Nagai for discussing the relation between TD-qHF and the semiconductor Bloch equation. This research was supported in part by Grants-in-Aid for Scientific Research (Nos. 18K14145, 19H02623, and 20H05670) from the Japan Society for the Promotion of Science (JSPS).
\end{acknowledgments}

\appendix

\appendix

\section{Keldysh parameter from a jump time in the Landau-Zener transition}
\label{section:appendixKeldysh}
In the analysis of the LZ model by Vitanov \cite{Vitanov}, {\it jump time} is introduced as a characteristic time where the diabatic population is divided by the slope of the population such that the diabatic energies are crossed. The jump time (Eq. (21) in \onlinecite{Vitanov}), when written as our variables, is given by 
\begin{align}
 T^{\mathrm{jump}} = \frac{\Delta}{\frac{1}{2}(G-G')E_{\mathrm{DC}}}
 =
 2 \frac{\sqrt{\mu \Delta}}{E_{\mathrm{DC}}}.
\end{align}
This formula is physically interpreted as the inverse of energy associated with the field strength $E_{\mathrm{DC}}$ times the penetration depth $1/\sqrt{\mu \Delta}$ in the energy gap $\Delta$ between the valence and conduction bands. This equation can be obtained by taking the ratio between the diagonal and off-diagonal components of our equation of motion \eqref{eq:EoM-LZ}. The Keldysh parameter in \onlinecite{Keldysh} with an angular frequency $\omega$ is constituted by
\begin{align} 
 \gamma_{\mathrm{Keldysh}} = \omega \frac{T^{\mathrm{jump}}}{2}.
\end{align}
The Keldysh parameter comes from the Keldysh theory for ionization in semiconductors via strong electric field\cite{Keldysh}.

\section{Density-matrix expression of TD-qHF and Houston basis representation}
\label{section:appendixB}
We introduce an alternative expression for TD-qHF with a density-matrix (DM)-based expression.
This expression is formally elegant and simple compared to the orbital expression introduced in Sect. \ref{section:theoretial_framework}.
We can find an explicit connection between SBE and the Wannier equation derived from TD-qHF.
Furthermore, this density-matrix-based equation can include phenomenological many-body influences such as a scattering term with relaxation time approximation.
Therefore, we derive here a generic theoretical framework in which the Bloch orbital is not used, since DM composed by a Bloch orbital is not necessarily justified when we add a phenological treatment in the equation of motion beyond the qHF wave function.
We only impose BvK boundary condition, $\rho(x+N_k a, x')=\rho(x,x'+N_k a)=\rho(x,x')$, and the lattice periodicity for the simultaneous spatial translation, $\rho(x+a,x'+a) = \rho(x,x')$.
Throughout this section, $t$ in the variables is omitted except for parametrically time-dependent ones.
In other words, the dynamical degrees of freedom to be determined by the equation of motion do not have explicit time coordinates in the following equations.

The total energy as a functional of DM is given as
\begin{align}
 E[\rho](t)
 =
 \iint\! \mathrm{d}x \mathrm{d}x' h(x',x;t)\rho(x,x')
 +
 E_{\mathrm{MF}}[\rho],\\
 E_{\mathrm{MF}}[\rho]
 =
 \frac{1}{2}\iint \! \mathrm{d}x \mathrm{d}x' \left[\rho(x,x)-\rho_0(x,x;t)\right]w(x,x')\left[\rho(x,x)-\rho_0(x,x;t)\right] \\
 -
 \frac{1}{4}\iint \! \mathrm{d}x \mathrm{d}x' \left[\rho(x,x')-\rho_0(x,x';t)\right]w(x',x)\left[\rho(x',x)-\rho_0(x',x;t)\right],\\
 h(x,x';t)
 =
 \delta(x-x')\left[ \left( -\mathrm{i}\frac{\partial}{\partial x'}+A(t)\right)^2 + v(x)\right], 
\end{align}
where spatial integration is taken over the simulation cell $0 \leq x,x' < N_k a$. We define a functional derivative of $F$ with respect to DM as
\begin{align}
 \left. \frac{\delta F}{\delta\rho (x,x')} \right|_{\rho}
 =
 \lim_{\epsilon \to 0}
 \frac{F[\rho (x,x') + \epsilon \delta \rho(x,x')] - F[\rho(x,x')]}{\epsilon} \label{eq:FunctionalDerivativewrtDM}.
\end{align}
Physically proper conditions for DM are not taken into account when this functional derivative is taken.
In other words, $\rho + \epsilon \delta \rho$ can be no physical DM even if $\rho$ satisfied physically expected conditions, such as $N$-representability for Fermionic one-body reduced DM. 
We introduce one-body Hamiltonian $h_{\mathrm{qHF}}$ by a functional derivative of the energy as 
\begin{align} 
 h_{\mathrm{qHF}}(x',x;t) = \frac{\delta E(t)}{\delta\rho (x,x')}
 =
 h(x',x;t) + v_{\mathrm{MF}}(x',x)\notag \\
 v_{\mathrm{MF}}(x',x)
 =
 \left( \int \! \mathrm{d}x'' w(x,x'') [\rho(x'', x'')-\rho_0(x'', x'';t)]\right) \delta(x'-x) \notag \\
 -
 \frac{1}{2} w(x',x) \left[ \rho(x',x) - \rho_0(x',x;t)\right].
\end{align}
We can then derive the equation of motion for DM from Eq. \eqref{eq:TD-qHF} as 
\begin{align} 
 \mathrm{i} \frac{\partial}{\partial t}\rho(x,x') \notag \\
 =
 \int \! \mathrm{d}x'' \left[h_{\mathrm{qHF}}(x,x'';t) \rho(x'',x') - \rho(x,x'') h_{\mathrm{qHF}}(x'',x';t) \right]. \label{eq:EoM-DM-TD-qHF}
\end{align}
The total energy time derivative is given as 
\begin{align} 
 \frac{\mathrm{d}}{\mathrm{d}t}E
 =
 \iint \! \mathrm{d}x\mathrm{d}x' \frac{\delta E}{\delta\rho (x,x')} \frac{\partial}{\partial t}\rho(x,x')
 +
 \mathrm{Tr}\left( \frac{\partial h(t)}{\partial t}\rho \right) \notag \\
 +
 \left( \frac{\partial \rho_0(t)}{\partial t} \textrm{ related terms} \right) \notag \\
 =
 -\mathrm{i}\mathrm{Tr} \left( h_{\mathrm{qHF}} h_{\mathrm{qHF}} \rho\right)
 +
 \mathrm{i}\mathrm{Tr} \left( h_{\mathrm{qHF}} \rho h_{\mathrm{qHF}}\right) \notag \\
 +
 \frac{\mathrm{d}A}{\mathrm{d}t} \mathrm{Tr}\left( \left[ -\mathrm{i}\frac{\partial}{\partial x}+A(t)\right]\rho \right) 
 -
 \frac{1}{2} \mathrm{Tr}\left( \left[ \frac{\partial \rho_0(t)}{\partial t} w \right]\rho \right) ,
\end{align}
where the trace is achieved by the spatial integration over the simulation cell. To obtain the last line, we use $\partial \rho_0(t)/\partial t = \mathrm{i}\dot{A} [x, \rho_0(t)]$ and $\mathrm{Tr} \left( \left[ x, \rho_0(t)\right] \rho_0(t)\right) =0$. The first and second terms in the final equation cancel each other out by the cyclic exchange in the trace. Therefore, the total energy is kept constant when the $A$-field is a constant in time.

We derive an expression to solve \eqref{eq:EoM-DM-TD-qHF} in the Houston basis \cite{Houston1940} without a time-dependent phase factor as 
\begin{align} 
 \rho(x,x') = \frac{2}{N_k} \sum_{\alpha \beta k} e^{\mathrm{i}kx} v_{\alpha, k+A(t)} (x) \rho_{\alpha\beta}^k v_{\beta, k+A(t)}^* e^{-\mathrm{i}kx'} \label{eq:Houstonbasi_for_DM},
\end{align}
where justification of the $k$-dependent decompositions is presented in Appendix \ref{section:NLoperator_formula}. The reason for phase factor omission here is to resemble SBE for TD-qHF. Matrix elements for a spatially nonlocal function $O(x,x')$ of the basis are defined as 
\begin{align}
 O_{\alpha \beta}^k (t)
 =
 \frac{1}{N_k} \int_0^{N_k a} \! \mathrm{d}x \int_0^{N_k a} \! \mathrm{d}x' \ e^{-\mathrm{i}kx} v_{\alpha, k+A(t)}^* (x) O(x,x';t) e^{+\mathrm{i}kx'} v_{\beta, k+A(t)}(x').
\end{align}
Substituting Eq. \eqref{eq:Houstonbasi_for_DM} into \eqref{eq:EoM-DM-TD-qHF}, we obtain the equation of motion for $\rho_{\alpha \beta}$ as
\begin{align} 
 \mathrm{i} \frac{\mathrm{d}}{\mathrm{d}t}\rho^k_{\alpha \beta} 
 =
 \left[ \epsilon_{\alpha k+A(t)} - \epsilon_{\beta k+A(t)} \right] \rho_{\alpha \beta}^k \notag \\
 +
 \sum_\gamma \Omega_{\alpha \gamma}^k(t) \rho_{\gamma\beta}^k
 -
 \sum_\gamma \rho_{\alpha\gamma}^k \Omega_{\gamma \beta}^k(t), \label{eq:EoM-DM-quasiHouston-TD-qHF} \\
 \Omega_{\alpha \beta}^k(t)
 =
 -\dot{A}(t) \mathcal{X}_{\alpha \beta}(k+A(t)) + v_{\mathrm{MF},\alpha\beta}^{k}, \notag \\
 \mathcal{X}_{\alpha \beta}(k)
 =
 \left\langle v_{\alpha k} \left| \frac{\partial v_{\beta k}}{\partial k} \right. \right\rangle
 =
 \int_0^a \! \mathrm{d}x v_{\alpha k}^* \frac{\partial v_{\beta k}(x)}{\partial k}, \label{eq:Generalized_Rabi_frequency_TD-qHF}
\end{align}
where $\left\langle v_{\alpha k} \left| \frac{\partial v_{\beta k}}{\partial k} \right. \right\rangle = -\left\langle \left. \frac{\partial v_{\alpha k}}{\partial k} \right| v_{\beta k} \right\rangle$ is used. $\Omega_{\alpha\beta}^k$ is the generalized Rabi frequency for TD-qHF. Explicit construction of the MF-term matrix element for $\varrho = \rho, \rho_0$ is given as
\begin{align}
 v_{\mathrm{MF}, \alpha \beta}^k [\varrho]
 = \notag \\
 \sum_{H} w^{q=0}(H) \eta(H)\left\langle v_{\alpha k+A(t)} \left| e^{\mathrm{i}Hx} \right| v_{\beta k+A(t)} \right \rangle \notag \times \\
 \left( \sum_{k' \gamma\delta} \varrho_{\gamma \delta}^{k'} \left\langle v_{\gamma k+A(t)} \left| e^{-\mathrm{i}Hx} \right| v_{\delta k+A(t)} \right \rangle \right) \notag \\
 -
 \frac{1}{2}\sum_{qH \gamma \delta} w^{q}(H) \varrho^{k-q}_{\gamma \delta}
 \left\langle v_{\alpha, k+A(t)} \left| e^{\mathrm{i}Hx} \right| v_{\gamma ,k-q+A(t)} (t) \right\rangle \times \notag \\
 \left\langle v_{\delta, k-q+A(t)} (t) \left| e^{-\mathrm{i}Hx} \right| v_{\beta ,k+A(t)} \right \rangle.
\end{align}

We introduce two approximations: first, that the Coulombic interaction has only a long-range part,
\begin{align}
 w^{q}(H) = \bar{w}^q \delta_{H,0},
\end{align}
and second, that the LPP labeled by the eigenfunction is the same for all crystal momenta,
\begin{align}
 \left\langle \left. v_{\alpha k} \right| v_{\alpha k'}\right\rangle = 1.
\end{align}
Within the approximation, we obtain the mean-field part as
\begin{align}
 v_{\mathrm{SBE},\alpha\beta}^k [\rho - \rho_0]
 =
 -
 \frac{1}{2}\sum_{q} \bar{w}^{q} \left\{ \rho_{\alpha \beta}^{k-q}(t) - \rho_{0,\alpha \beta}^{k-q} \right\}.
\end{align}
The equation of motion with the approximation is 
\begin{align}
 \mathrm{i} \frac{\mathrm{d}}{\mathrm{d}t}\rho^k_{\alpha \beta} 
 =
 \left[ \epsilon_{\alpha k+A(t)} - \epsilon_{\beta k+A(t)} \right] \rho_{\alpha \beta}^k \notag \\
 +
 \sum_\gamma \Omega_{\mathrm{SBE}, \alpha \gamma}^k(t) \rho_{\gamma\beta}^k
 -
 \sum_\gamma \rho_{\alpha\gamma}^k \Omega_{\mathrm{SBE}, \gamma \beta}^k(t), \label{eq:EoM-DM-SBE} \\
 \Omega_{\mathrm{SBE},\alpha \beta}^k(t)
 =
 -\dot{A}(t) \mathcal{X}_{\alpha \beta}(k+A(t)) + v_{\mathrm{SBE},\alpha\beta}^k.
\end{align}

We restrict ourselves to a two-band system achieved by $(\alpha,\beta)=(v,v), (v,c), (c,v), (c,c)$. The equations of motion are given as 
\begin{align}
 \mathrm{i} \frac{\mathrm{d}}{\mathrm{d} t}\rho_{vc}^k 
 =
 \left[ e_{c k}(t) - e_{v k}(t) \right] \rho_{vc}^k \notag \\
 +
 \left[ \rho_{vv}^k (t) - \rho_{cc}^k (t) \right]\Omega_{\mathrm{SBE}, vc}^k [\rho ],\notag \\
 e_{c k}(t)
 =
 \epsilon_{c k+A(t)} 
 -
 \frac{1}{2}\sum_{q} \bar{w}^{q} \rho_{cc}^{k-q}, \notag \\
 e_{v k}(t)
 =
 \epsilon_{v k+A(t)} 
 -
 \frac{1}{2}\sum_{q} \bar{w}^{q} \left( \rho_{vv}^{k-q} - 2 \right), \notag \\
 \mathrm{i} \frac{\mathrm{d}}{\mathrm{d} t}\rho_{vv}^k 
 =
 \rho_{v c}^k \Omega_{\mathrm{SBE}, cv}^k [\rho ] - \Omega_{\mathrm{SBE}, vc}^k [\rho ] \rho_{c v}^k , \notag \\
 \mathrm{i} \frac{\mathrm{d}}{\mathrm{d} t}\rho_{cc}^k (t)
 =
 \rho_{cv}^k \Omega_{\mathrm{SBE},vc}^k [\rho ] - \Omega_{\mathrm{SBE},cv}^k [\rho ] \rho_{vc}^k \label{eq:SBE}.
\end{align}
These are nothing but SBE, while minor differences are as follows.
In the original SBE construction, a bare electronic structure is given as a reference system, and it is renormalized by the electron-hole interaction even without an external field.
In our TD-qHF formulation, an already renormalized electronic structure is introduced by subtracting the initial density matrix $\rho_0$.
In front of the electron-hole attraction, the $1/2$ factor appears because of the exchange interaction for the spin-restricted electron system within qHF.
The intraband motion in the eigenvalue $\epsilon_{v k+A(t)}$ comes from the velocity gauge ansatz for the equation of motion.
In the velocity gauge, the counterpart to the dipole operator matrix element is $\mathcal{X}_{\alpha \beta}$ by taking into account $\dot{A} = -E(t)$.

Within two-band SBE, the gap renormalization of relative energy between conduction and valence bands is calculated as
\begin{align}
 e_{c k}(t) - e_{v k}(t)
 =
 \epsilon_{c k+A(t)} - \epsilon_{v k+A(t)} - \sum_{q} \bar{w}^{q} \rho_{cc}^{k-q} \label{eq:gap-renormalization_SBE},
\end{align}
where a conservation rule $\rho_{cc}^{k-q}(t) + \rho_{vv}^{k-q}(t) = 2$ is used. Thus, the band gap shrinks by increasing the conduction band population.

For the Wannier equation in our framework, we apply the linear perturbation theory for Eq. \eqref{eq:SBE} by $\rho(t) \simeq \rho_0 + \delta \rho(t)$ and neglect the momentum shift due to the $A$-field. The equation for $\delta \rho$ is
\begin{align}
 \mathrm{i} \frac{\mathrm{d}}{\mathrm{d} t} \delta \rho_{vc}^k 
 =
 \left[ \epsilon_{c k} - \epsilon_{v k} \right] \delta \rho_{vc}^k 
 -
 \sum_{q} \bar{w}^{q} \delta \rho_{vc}^{k-q}
 +
 2 \mathcal{X}_{vc}^k E(t). \label{eq:TDWannier-FS}
\end{align}
By taking the correspondence $k \to \hat{k} =-\mathrm{i}\partial /\partial X$, we obtain the real-space counterpart via Fourier transformation:
\begin{align}
 \mathrm{i} \frac{\mathrm{d}}{\mathrm{d} t} W_{vc} (X, t)
 =
 \left[ \epsilon_{c} ( \hat{k}) - \epsilon_{v} (\hat{k}) - \bar{w}(X) \right] W_{vc} (X,t) 
 +
 2 \mathcal{X}_{vc}(X) E(t). \label{eq:TDWannier-RS}
\end{align}
When the eigenvalue difference is approximated as $\epsilon_{c k} - \epsilon_{v k} \simeq \Delta + \frac{k^2}{2\mu}$, an eigenvalue equation for the homogneous part of \eqref{eq:TDWannier-RS} is
\begin{align}
  \left[ \Delta - \frac{1}{2\mu} \frac{\partial^2}{\partial X^2} - \bar{w}(X) \right] W_{vc} (X)
  =
  E_{\mathrm{op}} W_{vc} (X)  \label{eq:Wannier-RS}
\end{align}
This is the Wannier equation.

If we use the Houston basis with the trivial phase factor, as
\begin{align} 
 \rho^k(x,x') = \sum_{\alpha \beta} e^{-\mathrm{i}\int_0^t \! \mathrm{d}t' \epsilon_{\alpha k+A(t')} } v_{\alpha, k+A(t)} \tilde{\rho}_{\alpha\beta}^k v_{\beta, k+A(t)}^* e^{+\mathrm{i}\int_0^t \! \mathrm{d}t' \epsilon_{\beta k+A(t')} },
\end{align}
where $(v_{\alpha, k}, \epsilon_{\alpha, k})$ is the eigenpair of the field-free Hamiltonian without an MF part, $\int \! \mathrm{d}x' h(x,x';t=0) e^{\mathrm{i}kx' } v_{\alpha, k}(x') =\epsilon_{\alpha k}e^{\mathrm{i}kx } v_{\alpha, k}(x) $. The equation of motion for $\rho_{\alpha\beta}^k$ reads
\begin{align} 
 \mathrm{i} \frac{\mathrm{d}}{\mathrm{d}t}\tilde{\rho}^k_{\alpha \beta} 
 =
 \sum_\gamma e^{+\mathrm{i}\int_0^t (\epsilon_{\alpha k+A(t')} - \epsilon_{\gamma k+A(t')})} \Omega_{\alpha \gamma}^k(t) \tilde{\rho}_{\gamma\beta}^k\notag \\
 -
 \sum_\gamma e^{-\mathrm{i}\int_0^t (\epsilon_{\beta k+A(t')} - \epsilon_{\gamma k+A(t')})} \tilde{\rho}_{\alpha\gamma}^k \Omega_{\gamma \beta}^k(t), \label{eq:EoM-DM-Houston-TD-qHF} 
\end{align}
where the definition of the generalized Rabi frequency $\Omega_{\gamma \beta}^k(t)$ is the same as \eqref{eq:Generalized_Rabi_frequency_TD-qHF}.
This equation of motion is simply another expression of Eq. (9)) in \onlinecite{Ikemachi2018}.
$\tilde{\rho}_{\alpha\beta}$ varies slowly compared to the energy difference $\epsilon_{\beta k}-\epsilon_{\alpha k}$ because of the phase factor inclusion in the Houston basis representation.
The term that has $-\dot{A}(t)\mathcal{X}$ is expected to be off-resonant because $\left\langle v_{\alpha k} \left| \partial v_{\gamma k}/\partial k \right. \right\rangle_{k+A(t)}$ has only slowly varying components.
Resonant oscillation component only comes from $v_{\mathrm{MF},\alpha \gamma}^k\simeq e^{-\mathrm{i}\int_0^t (\epsilon_{\alpha k+A(t')} - \epsilon_{\gamma k+A(t')})}$ as pointed out by \onlinecite{Ikemachi2018}.

\section{Nonlocal function with lattice periodicity}
\label{section:NLoperator_formula}
Generally, a function with two spatial coordinates $f(x,x')$ requires two wave numbers for the Fourier transformation.
Here, we prove a function that satisfies $f(x+N_k a, x') = f(x,x'+N_k a) = f(x,x'), f(x+a,x'+a) = f(x,x')$ has a particular form for the Fourier transformation that has single crystal momentum and two reciprocal lattice coordinates.
We prove this form to validate decomposition form \eqref{eq:Houstonbasi_for_DM}.

Consider a function $f(x,x')$ within the BvK boundary condition $f(x+N_k a, x') = f(x,x'+N_k a) = f(x,x')$. We further assume a lattice periodicity for simultaneous spatial variable translation as
\begin{align} 
 f(x+a, x'+a) = f(x,x').
\end{align}
This condition appears in the one-body reduced DM and one-body Green's function in typical many-body problems for perfect crystals.

This function can be expanded as
\begin{align} 
 f(x,x') = \sum_k e^{\mathrm{i}kx} f^k(x,x') e^{-\mathrm{i}kx}, \notag \\
 f^k(x+a,x') = f^k(x,x'+a) = f^k(x,x'),
\end{align}
where $k$ is the discretized wave-number $k=0, 2\pi /(N_k a), \dots, 2\pi /(N_k a)(N_k - 1)$, as in the orbital case.

This single $k$ expansion is proved by starting from a general Fourier expansion:
\begin{align} 
 f(x,x') = \sum_{kk'GG'} e^{+\mathrm{i}(k+G)x} \tilde{f}^{k,k'}(G,G') e^{-\mathrm{i}(k'+G')x'}, \notag \\
 \tilde{f}^{k,k'}(G,G')
 =
 \frac{1}{(N_k a)^2}\iint \! \mathrm{d}x \mathrm{d}x' e^{-\mathrm{i}(k+G)x} f(x,x') e^{+\mathrm{i}(k'+G')x'},
\end{align}
where $k,k'$ are discretized Brillouin zone indices and $G,G'$ varies the reciprocal lattices. Using the simultaneous translation for this formula, we obtain 
\begin{align} 
 f(x+a,x'+a) = \sum_{kk'GG'} e^{+\mathrm{i}(k-k')a} e^{+\mathrm{i}(k+G)x} \tilde{f}^{k,k'}(G,G') e^{-\mathrm{i}(k'+G')x'}.
\end{align}
Since all components are the same, we obtain $e^{+\mathrm{i}(k-k')a} = 1$ for an arbitrary $(k,k')$ pair. This means that $k'$ is equal to $k$, namely, a single $k$-index is sufficient for $\tilde{f}^{k,k'}(G,G) \to f^{k}(G,G')$. Thus, the proper expansion is given as 
\begin{align} 
 f(x,x') = \sum_{kGG'} e^{+\mathrm{i}(k+G)x} f^{k}(G,G') e^{-\mathrm{i}(k+G')x'}, \notag \\
 f^{k}(G,G')
 =
 \frac{1}{(N_k a)^2}\iint \! \mathrm{d}x \mathrm{d}x' e^{-\mathrm{i}(k+G)x} f(x,x') e^{+\mathrm{i}(k+G')x'}.
\end{align}
By taking the partial sum over the reciprocal lattices, we have the following formula:
\begin{align} 
 f(x,x') = \sum_{k} e^{+\mathrm{i}kx} f^{k}(x,x'') e^{-\mathrm{i}kx'}, \notag \\
 f^{k}(x,x')
 =
 \sum_{GG'} e^{+\mathrm{i}Gx} f^{k}(G,G') e^{-\mathrm{i}G'x'}.
\end{align}
$f^k(x+a,x') = f^k(x,x'+a) = f^k(x,x')$ can be confirmed from this formula.

Note that we do not impose any condition on $f$ except for the BvK boundary condition and the simultaneous lattice translation symmetry. We can derive a more specific formula with a more explicit form for $f$. One-body reduced DM composed by a Slater determinant with Bloch orbitals, for example, has the following factorized shape:
\begin{align}
 \rho(x,x') = \frac{2}{N_k}\sum_{ik} e^{+\mathrm{i}kx} u_{ik}(x) u_{ik}^*(x') e^{-\mathrm{i}kx'}.
\end{align}
The Fourier component of $k$-dependent DM also has a factorial shape:
\begin{align}
 \rho^k(G,G') = \frac{2}{N_k}\sum_{i} u_{ik}(G) u_{ik}^*(G').
\end{align}

\end{document}